\documentclass[]{jfm}

\usepackage{graphicx}
\usepackage{newtxtext}
\usepackage{newtxmath}
\usepackage{natbib}
\usepackage[hidelinks]{hyperref}
\hypersetup{
    colorlinks = true, 
    urlcolor   = blue,
    citecolor  = blue
}
\usepackage[justification=justified,
   format=plain,width=\textwidth]{caption}

\newcommand{\RomanNumeralCaps}[1]

\graphicspath{{figures/}}
\usepackage{amsmath}
\usepackage[inkscapeformat=png]{svg}
\usepackage{calrsfs}

\DeclareMathOperator{\ra}{\mathit{Ra}}

\usepackage{color}

\title[Solute mixing in porous media with dispersion and buoyancy]{Solute mixing in porous media with dispersion and buoyancy}

\author{
Marco De Paoli\aff{1,2}\corresp{\email{m.depaoli@utwente.nl ; marco.de.paoli@tuwien.ac.at}},
Guru Sreevanshu Yerragolam\aff{1},
Roberto Verzicco\aff{1,3,4},
\and
Detlef Lohse\aff{1,5}
}
\shortauthor{De Paoli M., Yerragolam G.S., Verzicco R. \& Lohse D.}

\affiliation{
\aff{1}Physics of Fluids Department and Max Planck Center for Complex Fluid Dynamics and J. M. Burgers Centre for Fluid Dynamics, University of Twente, P.O. Box 217 7500AE Enschede, The Netherlands
\aff{2}Institute of Fluid Mechanics and Heat Transfer, TU Wien, 1060 Vienna, Austria
\aff{3}Dipartimento di Ingegneria Industriale, University of Rome ``Tor Vergata'', 00133 Roma, Italy
\aff{4}Gran Sasso Science Institute, 67100 L'Aquila, Italy
\aff{5}Max Planck Institute for Dynamics and Self-Organization, Am Faßberg 17, Göttingen 37077, Germany
}

\begin{document}
\maketitle

\begin{abstract}
We analyse the process of convective mixing in two-dimensional, homogeneous and isotropic porous media with dispersion. We considered a Rayleigh-Taylor instability in which the presence of a solute produces density differences driving the flow. The effect of dispersion is modelled using an anisotropic Fickian dispersion tensor (Bear, \emph{J. Geophys. Res.} vol. 66, 1961, pp. 1185–1197). In addition to molecular diffusion ($D_m^*$), the solute is redistributed by an additional spreading, in longitudinal and transverse flow directions, which is quantified by the coefficients $D_l^*$ and $D_t^*$, respectively, and it is produced by the presence of the pores. The flow is controlled by three dimensionless parameters: the Rayleigh-Darcy number $\ra$, defining the relative strength of convection and diffusion, and the dispersion parameters $r=D_l^*/D_t^*$ and $\Delta=D_m^*/D_t^*$. With the aid of numerical Darcy simulations, we investigate the mixing dynamics without and with dispersion. We find that in absence of dispersion ($\Delta\to\infty$) the dynamics is self-similar and independent of $\ra$, and the flow evolves following several regimes, which we analyse. Then we analyse the effect of dispersion on the flow evolution for a fixed value of the Rayleigh-Darcy number ($\ra=10^4$). A detailed analysis of the molecular and dispersive components of the mean scalar dissipation reveals a complex interplay between flow structures and solute mixing. We find that the dispersion parameters $r$ and $\Delta$ affect the fingers formation and their dynamics: the lower the value of $\Delta$ (or the larger the value of $r$), the wider, more convoluted and diffused the fingers. We also find that for strong anisotropy, $r=O(10)$, the role of $\Delta$ is crucial: except for the intermediate phases of the flow dynamics, dispersive flows show more efficient (or at least comparable) mixing than in non-dispersive systems.
Finally, we look at the effect of the anisotropy ratio $r$, and we find that it produces only second order effects, with relevant changes limited to the intermediate phase of the flow evolution, where it appears that the mixing is more efficient for small values of anisotropy. The proposed theoretical framework, in combination with pore-scale simulations and bead packs experiments, can be used to validate and improve current dispersion models to obtain more reliable estimates of solute transport and spreading in buoyancy-driven subsurface flows.
\end{abstract}


\section{Introduction}\label{sec:rt_intro}
The transport and mixing of a solute or a dispersed phase within a fluid-saturated porous medium is a key process common to many subsurface flows, including petroleum migration \citep{simmons2001variable}, geological sequestrations of carbon dioxide \citep{Emami-Meybodi2015} and underground hydrogen storage \citep{krevor2023subsurface}.
These processes are controlled by a convective flow driven by the local fluid density differences, due to a non-uniform solute distribution within the domain.
The evolution of these systems is hard to monitor due to the inaccessibility of the sites, typically located hundreds of meters underground.
Nevertheless, an accurate prediction of solute transport in subsurface buoyancy-driven flows is required to address key environmental challenges, including also the design of subsurface storage sites for radioactive waste \citep{woods2015flow} and the development of remediation strategies for contaminated regions \citep{leblanc1984sewage,molen1988,bear2010modeling}.
For instance, to design the groundwater remediation strategies for light and dense non-aqueous petroleum liquids (LNAPL, DNAPL), resulting from spillage of fuels or chemicals, it is key to determine the area over which these contaminants spread, and then where the remediation should take place \citep{woods2015flow}.

Accurate predictions of the above mentioned flows are made further complex by the interplay of convective and diffusive processes, representing the driving and dissipative mechanics, respectively, of these systems \citep{depaoli2023review}, and therefore controlling mixing.
At the pore-scale and in absence of flow, solute transport is regulated by molecular diffusive mixing, acting to reduce local gradients of solute concentration: caused by the Brownian motion of the molecules, it produces a flux of solute from regions of high concentration towards regions of low concentration.
When the fluid moves within the intricate interstitial pore channels, the fluid and the solute cannot penetrate into the solid obstacles (e.g., rock grains) forming the porous matrix, and will follow random walk-type paths resulting in a further solute redistribution in the flow (longitudinal) direction \citep{woods2015flow}.
In addition, the velocity at the pore scale varies in magnitude and direction from point to point within the fluid present in the void space.
This causes fluid particles travelling along its own microscopic streamline to spread out, contributing to further disperse in the flow direction any solute contained in it.
The velocity gradient existing within the pores in the direction perpendicular (also defined transverse) to the flow, causes a differential transport of solute in that direction, leading to gradients of concentration, and ultimately producing solute fluxes in transverse direction due to molecular mixing. 
The mechanisms of spreading produced by a flow through a porous medium are referred to as ``mechanical'' dispersion, to indicate a spreading due to fluid mechanical phenomena.
Unlike molecular diffusion which occurs also in a still fluid, both the flow and the porous medium are required for mechanical dispersion to take place.
We refer to \citet{bear2010modeling} for an extensive presentation of this processes.
We remark that molecular diffusion is a pore-scale process purely controlled by concentration gradients.
Despite being originated by a very different physical mechanism, mechanical dispersion (which is a macroscopic flow property) has been modelled through a functional form analogue to that of molecular diffusion, since both contribute to spread the solute.
In addition, the effects of mechanical dispersion may overcome those of molecular diffusion in absence of flow by several orders of magnitude. 
At the macroscopic (Darcy) level, the combination of molecular diffusion and mechanical dispersion is also indicated as hydrodynamic dispersion \citep{bear2010modeling,wen2018rayleigh}.
For simplicity, we will refer to ``mechanical dispersion'' as ``dispersion''.
Some of the modelling approaches proposed to describe the spreading of a solute in a porous medium include modelling of the shear \citep{taylor1953dispersion}, of the no-slip condition at the solid boundaries \citep{saffman1959theory} and of the dead-end pores \citep{coats1964dead}.

Deriving a unified model reproducing all these effects is challenging, due to the large space of the governing parameters (depending on fluid, medium and flow properties) and the different nature of the physical mechanisms involved.
An approach commonly adopted consists of introducing a dispersion tensor in the transport equation governing the scalar phase (solute).
This tensor combines molecular diffusion (quantified by the diffusivity coefficient $D_m^*$) and dispersion, modelled as an equivalent (or effective) value of solute diffusivity. 
Two coefficients are used to describe the spreading of solute in longitudinal (i.e., aligned with the flow) and transverse (i.e., perpendicular to the flow) directions \citep{delgado2007longitudinal}, and are indicated with $D_l^*$ and $D_t^*$, respectively.
In general, these coefficients are obtained through laboratory experiments in constant displacement flows through uniform porous media, but analogue results can be also achieved using pore-network models \citep{bijeljic2004pore}. 
It has been shown that the dispersion coefficients depend on the relative importance of flow velocity to solute diffusivity (P\'eclet number), among other parameters (e.g., Schmidt number and surface tension), and several flow regimes have been identified and well characterized theoretically \citep{saffman1959theory,koch1985dispersion,puyguiraud2021pore}.

In the above mentioned works, the flow field is imposed a priori, and it is independent of the solute distribution within the system.
In case of buoyancy-driven flows, however, the solute field determines the flow, and the effects of dispersion have been quantified only for a few configurations.
\citet{menand2005dispersion} provided measurements of the longitudinal dispersion coefficient for two miscible fluid layers with different density, and driven by either a gravitationally stable or unstable linear displacement flow, and therefore with dispersion provided by both displacement- and buoyancy-induced flows.
They found that when buoyancy dominates, representing the case of interest in this work, dispersion controls the initial diffusive growth of the mixing zone and the formation and growth of the instabilities.
From a macroscopic perspective, buoyancy-driven flows have been investigated by \citet{hesse2010buoyant}.
They observed that the effect of dispersion associated with the presence of barriers is different from that relative to small-scale dispersion and, depending on the geometry of the barriers, may exceed the latter by several orders of magnitude.
The role of dispersion on solute transport in buoyancy-driven flows has been partially explored via numerical simulations in semi-infinite systems \citep{hidalgo2012scaling,emami2017dispersion,Liang2018,Michel-Meyer2020,dhar2022convective} and in confined steady-state flows with constant driving \citep{gasow2021macroscopic,wen2018rayleigh}.
Despite these efforts, the effects of dispersion on buoyancy-driven flows in closed systems (i.e., in absence of an external driving) remain largely unexplored, and we aim precisely at bridging this gap. 

We analyse the role of dispersion on convective mixing in two-dimensional, homogeneous and isotropic porous media.
We consider a Rayleigh-Taylor instability, obtained by stacking, in an unstable configuration, two fluid layers initially divided by a horizontal interface.
The fluids are fully miscible, so there is no stabilizing effect introduced by the presence of a surface tension, and a solute is responsible for the density differences which drive the flow.
This flow configuration represents the archetypal problem used to study closed systems, and has been extensively studied in previous numerical works, in two-dimensions and without dispersion \citep[see, e.g.,][and references therein.]{dewit2016chemo,dewit2020chemo,depaoli2019prf,gopalakrishnan2021scalings}
Recently this problem has been also investigated in detail in 3D \citep{Boffetta2020,boffetta2022dimensional}, allowing to explore the role of the flow dimensionality.
In this work, we include also the effect of dispersion of solute, which using the anisotropic Fickian dispersion tensor formulation proposed by \citet{bear1961tensor}.
In this modelling framework, dispersion is controlled by the dimensionless parameters $r=D_l^*/D_t^*$, defined as the anisotropy ratio, and by the relative importance of molecular diffusion to transverse dispersion, $\Delta=D_m^*/D_t^*$.
While $\Delta$ may vary over several orders of magnitude \citep[e.g., $0.02\le\Delta\le100$, see][]{Liang2018}, $r$ is restricted to a narrower range, namely $1\le r\le O(10)$, where $r\approx10$ is appropriate for advection-dominated flows with solute transport \citep{bijeljic2007pore}.
The relative strength of buoyancy and molecular diffusion is quantified by the Rayleigh-Darcy number ($\ra$).
With the aid of numerical Darcy simulations, we investigate the mixing dynamics at $\ra=10^4$, in correspondence of which multiple flow regimes are present (unlike at lower values of $\ra$).
We employ the anisotropic dispersion tensor formulation to simulate high-$\ra$ convection in a closed system.
We use the model proposed by \citet{wen2018rayleigh}, who investigated a statistically-steady porous Rayleigh-B\'enard system, to analyse a transient flow. 
Moreover, while in the Rayleigh-B\'enard case heat or mass transfer at the boundaries is possible, the problem investigated here is a closed system with no external forcing, where the driving is uniquely given by the initial available potential energy.
Finally, this work complements recent studies on Rayleigh-Taylor convection in porous media at high-Rayleigh-Darcy numbers \citep{depaoli2019universal,depaoli2019prf,Boffetta2020,boffetta2022dimensional} by systematically investigating the effect of dispersion.

The paper is organized as follows. 
In~\S\ref{sec:darcy} we present the flow configuration and the numerical method employed. 
The theoretical framework developed to carry out a detailed analysis of the molecular and dispersive components of the mean scalar dissipation is presented in~\S\ref{sec:valid}.
The flow evolution is first studied in absence of dispersion ($\Delta\to\infty$, \S\ref{sec:resndisp}), and then the effect of $\Delta$ (\S\ref{sec:delta}) and $r$ (\S\ref{sec:r}) are considered. 
Finally, the conclusions, an example of relevance to geophysical applications and limitations of the current approach are discussed in \S\ref{sec:concl}.

\section{Methodology}\label{sec:darcy}
We study a convective flow in a porous medium at the Darcy scale.
In this framework, the equations are written for quantities averaged over a Reference (or Representative) Elementary Volume (REV) \citep{whitaker1998method}, consisting of an intermediate scale larger than the individual pores, but smaller than the characteristic size of the macroscopic domain and of the flow structures.
A thorough description on how to select an appropriate size of the REV is discussed by \citet{whitaker1998method} and \cite{nield2006convection}.
In general, the results of the averaging procedure should be independent of the size of the REV's size.
For buoyancy-driven flows, this procedure can be applied when: (i) the dissipative mechanisms, as molecular diffusion and viscous dissipation, dominate over inertia \citep{depaoli2023review}; and (ii) the flow structures are large compared to the characteristic pore-size \citep{hewitt2020vigorous}.

\begin{figure}
    \centering
    \includegraphics[width=0.8\columnwidth]{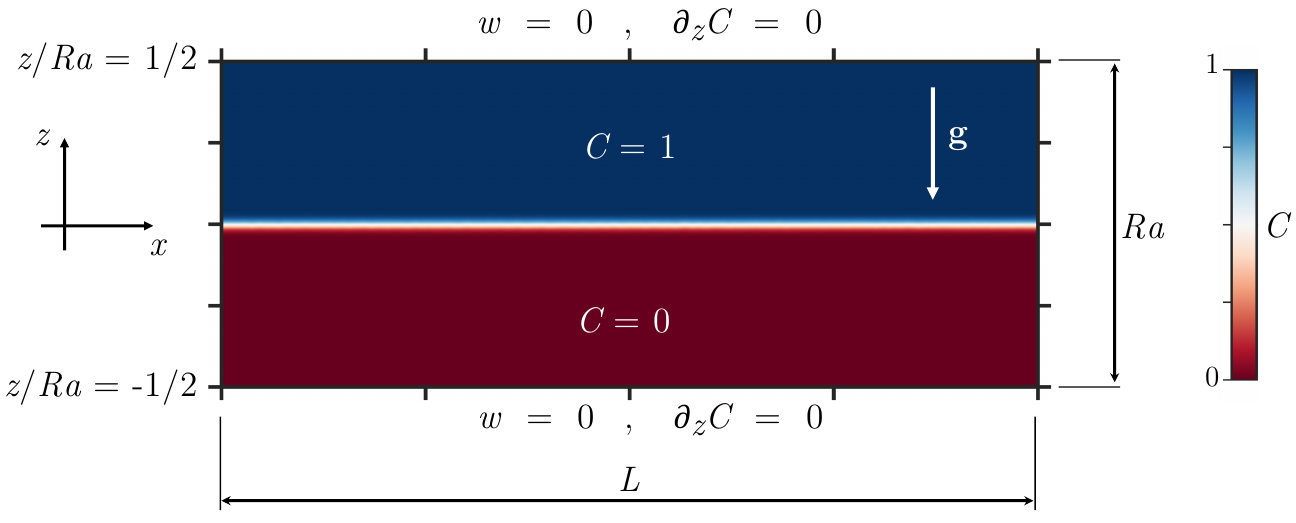}
    \caption{
    Sketch of the flow configuration with all quantities shown in dimensionless units. 
    An example of initial concentration field, $C$, consisting of a heavy fluid layer with maximum solute concentration at top ($C=1$) and minimum at bottom ($C=0$), is shown.    
    The flow reference frame ($x,z$), the boundary conditions and the direction of the gravity acceleration ($\mathbf{g}$) are indicated, as well as the domain size in horizontal ($L$) and vertical ($\ra$) directions.
    }
    \label{fig:intro0}
\end{figure}

We consider a fluid-saturated porous medium in a two-dimensional domain having uniform porosity $\phi$ and permeability $K$. 
We assume the flow is incompressible and governed by the Darcy equation, and it is characterized by an unstable density difference ($\Delta \rho^{*}$) induced by the presence of a solute concentration field, $C^*$ (the superscript $^{*}$ is used to indicate dimensional variables).
The system is periodic in horizontal direction $x^*$ and confined by two walls that are impermeable to fluid and solute in vertical direction $z^*$, along which the gravitational acceleration $\mathbf{g}$ acts.
The domain extension is $L^*_x$ and $L^*_z$ in horizontal and vertical directions, respectively. 
A sketch of the computational domain with indication of the boundary conditions is reported in figure~\ref{fig:intro0}.
The scalar field $C^{*}$ varies between $C^{*}_\text{min}$ and $C^{*}_\text{max}$.
The evolution of this field is controlled by the advection-dispersion equation \citep{nield2006convection}
\begin{equation}
\phi\frac{\partial C^{*}}{\partial t^{*}}+\mathbf{u}^{*}\cdot\nabla^{*}C^{*}=\nabla^* \cdot\left(\phi \mathsfbi{D}^* \nabla^{*} C^{*}\right) \text{  ,} 
\label{eq:eqadim3}
\end{equation}
where $t^{*}$ is time and $\mathbf{u}^{*}=(u^{*},w^{*})$ is the velocity field. 
The effect of dispersion is accounted by the hydrodynamic dispersion tensor $\mathsfbi{D}^*$ that depends on the local flow conditions.
We follow the formulation proposed by \citet{bear1961tensor} and later widely employed \citep{bear2012introduction,emami2017dispersion,wen2018rayleigh} to model dispersion effects, and $\mathsfbi{D}^*$ reads:
\begin{equation}
\mathsfbi{D}^* = D_m^*\mathsfbi{I}+(\alpha_l^*-\alpha_t^*)\frac{\mathbf{u}^*(\mathbf{u}^*)^{\textrm{T}}}{\lvert\mathbf{u}^* \rvert}+\alpha_t^*\mathsfbi{I}\lvert\mathbf{u}^* \rvert
\label{eq:disp01}
\end{equation}
where $\mathsfbi{I}$ is the identity matrix, $D_m^*$ the molecular diffusion coefficient, $\alpha_l^*$ the longitudinal dispersivity and $\alpha_t^*$ the transverse dispersivity \citep[where $\alpha_l^*\ge\alpha_t^*$,][]{delgado2007longitudinal}.
Note that the advection-diffusion form of the tensor $(\mathsfbi{D}^* = D_m^*\mathsfbi{I})$ is recovered when $\mathbf{u}^*\alpha_l^*/D_m^*\ll 1$.

We consider the fluid density, $\rho^{*}$, to be a linear function of the concentration:
\begin{equation}
\rho^{*}(C^{*})=\rho^{*}(C^{*}_\text{min})+\Delta\rho^{*}\frac{C^{*}-C^{*}_\text{min}}{\Delta C^*}
\text{  ,} 
\label{eq:eqadim4}
\end{equation}
with $\Delta C^* = C^{*}_\text{max}-C^{*}_\text{min}$ and $\Delta\rho^{*}=\rho^{*}(C^{*}_\text{max})-\rho^{*}(C^{*}_\text{min})$.
Assuming the validity of the Boussinesq approximation, which is reasonable, e.g., in the context of brine transport in porous media \citep{landman2007heat}, the flow field is fully described by the continuity and the Darcy equations,
\refstepcounter{equation}
$$
\nabla^{*}\cdot\mathbf{u}^{*}=0\quad,\quad
\mathbf{u}^{*}=-\frac{K}{\mu}\left(\nabla^{*} P^{*}+\rho^{*}g \mathbf{k}\right) \text{  ,} 
\eqno{(\theequation{\mathit{a},\mathit{b}})}\label{eq:eqadim2}
$$
with $\mu$ the fluid viscosity (constant), $P^{*}$ the pressure and $\mathbf{k}$ the vertical unit vector.
Additional correction terms to account for inertial effects can be included \citep{nield2006convection}, but lie beyond the scope of this work.
Since the walls are impermeable to the fluid, the boundary condition reads
\begin{equation}
\mathbf{u}^* \cdot \mathbf{n} = 0 \quad \Rightarrow \quad 
\begin{cases}
    w^*(z^*=-L^{*}_{z}/2)=0\\
    w^*(z^*=+L^{*}_{z}/2)=0   
\end{cases}
\label{eq:bcnopenet}
\end{equation}
with $\mathbf{n}$ the unit vector perpendicular to the boundary (note that slip at the walls is possible).
At the upper and lower walls, no flux ($\partial_{z^*} C^*=0$) conditions are considered.
Periodicity is forced in the wall-parallel directions.

\subsection{Dimensionless equations}\label{sec:dimless}
A natural velocity scale relevant to the convective system examined is the buoyancy velocity, $\mathcal{U}^{*}=g \Delta \rho^{*} K / \mu$.
After the onset of convection, fingers start to develop vertically from the centreline ($z^*=0$), and the domain height is relevant only after the fingers reach the walls.
Therefore, it may be convenient to make the equations dimensionless with respect to flow units that are independent of the domain geometry.
In particular, as proposed by \citet{fu2013pattern}, one can use as a reference length scale $\ell^*=\phi D_m^*/\mathcal{U}^*$, where the vertical domain extension $L^*_z$ is not included.
Using these scales we obtain the following set of dimensionless variables:
\begin{equation}
C=\frac{C^{*}-C^{*}_\text{min}}{C^{*}_\text{max}-C^{*}_\text{min}},\quad x=\frac{x^{*}}{\ell^{*}},\quad \mathbf{u}=\frac{\mathbf{u}^{*}}{\mathcal{U}^{*}},
\label{eq:eqadim5aaa}
\end{equation}
\begin{equation}
t=\frac{t^{*}}{\phi \ell^{*}/\mathcal{U}^{*}},\quad p=\frac{p^{*}}{\Delta \rho^{*}g\ell^{*}} ,
\label{eq:eqadim5}
\end{equation}
where we introduced the reduced pressure $p^{*}=P^*+\rho^*(C^*_\text{min})gz^*$, we finally derive the dimensionless form of the governing equations~\eqref{eq:eqadim3}-\eqref{eq:eqadim2}:
\begin{equation}
\label{eq:equ1bis1}
\frac{\partial C}{\partial t}+\mathbf{u}\cdot\nabla C = \nabla \cdot \left( \mathsfbi{D} \nabla C\right) 
\end{equation}
\begin{equation}
\nabla\cdot\mathbf{u}=0,
\label{eq:equ1bis2}
\end{equation}
\begin{equation}
\mathbf{u}=-\left( \nabla p + C \mathbf{k}\right) ,
\label{eq:equ1bis3}
\end{equation}
where  
\begin{equation}
\ra=\frac{g \Delta \rho^{*} K L^{*}_{z} }{ \phi D_m^* \mu}=\frac{\mathcal{U}^{*} L^{*}_{z}}{\phi D_m^*}
\label{eq:rada}
\end{equation}
is the Rayleigh-Darcy number (indicated as Rayleigh number in the following).
We recall here that $K$ and $\phi$ represent the medium permeability and porosity, respectively, and are considered uniform here. 
Finally, the dispersion tensor $\mathsfbi{D}^*$ introduced in~\eqref{eq:disp01} and expressed in dimensionless form reads:
\begin{equation}
\mathsfbi{D}=\mathsfbi{I}+\frac{1}{\Delta}\left[(r-1)\frac{\mathbf{u}\mathbf{u}^{\textrm{T}}}{\lvert\mathbf{u}\rvert}+\mathsfbi{I}\lvert\mathbf{u}\rvert\right] ,
\label{eq:disp01ad}
\end{equation}
where 
\begin{equation}
\Delta=\frac{D_m^*}{D_t^*}\quad,\quad r=\frac{D_l^*}{D_t^*}=\frac{\alpha_l^*}{\alpha_t^*}
\label{eq:params}
\end{equation}
with $D_m^*$ the molecular diffusion coefficient, $D_t^*=\alpha_t^*\mathcal{U}^*$ the transverse dispersion coefficient, $D_l^*=\alpha_l^*\mathcal{U}^*$ the longitudinal dispersion coefficient and $r$ the dispersivity ratio. 
The flow is completely defined by four dimensionless parameters: $\ra$, $L=L^*_x/\ell^*$, $\Delta$ and $r$.
This formulation is similar to that proposed by \citet{wen2018rayleigh}. However, in this case the governing parameter $\ra$ does not appear explicitly in the equations, and it corresponds to the dimensionless domain height ($\ra = L^{*}_{z} /\ell^*$, see also figure~\ref{fig:intro0}).
In addition, provided $L$ is sufficiently large to allow the formation of multiple flow structures and minimize the effects of the periodic forcing, the flow is also independent of $L$ itself. 

\subsection{Numerical solution of the equations}
The set of equations~\eqref{eq:equ1bis1}-\eqref{eq:equ1bis3} is solved numerically with the aid of the second-order finite difference code AFiD-Darcy open-sourced by our research group \citep{depaoli2025code}.
The code consists of an efficient solver for massively-parallel simulations of convective, wall-bounded and incompressible porous media flows, and it is based on the initial version of AFiD developed for turbulent flows \citep{van2015pencil}.
The algorithm is based on a pressure-correction scheme and employs an efficient Fast Fourier Transform-based solver.
The parallelization method is implemented in a two-dimensional pencil-like domain decomposition, which enables efficient parallel large-scale simulations.
The implementation in absence of dispersion was validated against different canonical flows in porous media, including Rayleigh-Taylor instability \citep{Boffetta2020}. 
Additional numerical details on the solution algorithm are provided by \citet{depaoli2025afid}.
In addition, in this work we solve the flow including the effect of dispersion~\eqref{eq:disp01ad} (see Appendix~\ref{sec:appA1} for further details on the numerical treatment of the dispersive terms).
The algorithm with dispersion is validated against the work of \citet{wen2018rayleigh} repeating their simulations in Rayleigh-B\'enard-Darcy configuration, and comparing the results in terms of molecular, dispersive and total Nusselt numbers.

We employ here a uniform grid spacing with cells of size $\Delta x\approx\Delta z<10$ (expressed in dimensionless units as discussed in \S\ref{sec:dimless}), in horizontal and vertical directions, respectively.
Present simulations are over-resolved compared to the minimum requirements of the case without dispersion \citep[corresponding to $\Delta x = \Delta z = 15.625$, see][]{depaoli2025afid}.
The minimum number of grid points in vertical direction used here is 128, making the grid even further over-resolved in the low-$\ra$ case.
In the horizontal direction, the domain size is fixed ($L=10^5$), as well as its resolution ($N_x=10240$, $\Delta x = L/N_x<10$), for all simulations except for the smallest $\Delta$ considered here ($\Delta=5\times10^{-2}$), for which the resolution is doubled and the domain width halved in order to keep the computational cost affordable. 
The details of all simulations performed are listed in table~\ref{tab:resdisp}. 
The flow is initialized with a step-like profile obtained from the analytical diffusive solution~\eqref{eq:equ1bis1}.
Additional details on the initial condition and the grid requirements are also provided in Appendix~\ref{sec:appA2}.

\begin{table}%
\centering
\begin{tabular}{c c c c c c c c c c c c} 
$\ra$ && $\Delta$ && $r$ && $L$ && $L/\ra$ & $N_x \times N_z$ && $\gamma$\\ \\
$1\times10^2$ && $\infty$ && $-$ && $1\times10^5$ && 1000 & $10240 \times 128 $&& \\
$2\times10^2$ && $\infty$ && $-$ && $1\times10^5$ && 500 & $10240 \times  128 $&& \\
$5\times10^2$ && $\infty$ && $-$ && $1\times10^5$ && 200 & $10240 \times  128 $&& \\
$1\times10^3$ && $\infty$ && $-$ && $1\times10^5$ && 100 & $10240 \times  128 $&& \\
$2\times10^3$ && $\infty$ && $-$ && $1\times10^5$ && 50 & $10240 \times 256 $&& \\
$5\times10^3$ && $\infty$ && $-$ && $1\times10^5$ && 20 & $10240 \times  512 $&& \\ 
$1\times10^4$ && $\infty$ && $-$ && $1\times10^5$ && 10 & $10240 \times  1024 $&& 0.59\\
$2\times10^4$ && $\infty$ && $-$ && $1\times10^5$ && 5 & $10240 \times  2048 $&& \\ \\
$1\times10^4$ && $1\times10^{5}$ && $10$ && $1\times10^5$ && 10 & $10240 \times  1024 $&& 0.59\\
$1\times10^4$ && $1\times10^{1}$ && $10$ && $1\times10^5$ && 10 & $10240 \times  1024 $&& 0.59\\
$1\times10^4$ && $1\times10^{0}$ && $10$ && $1\times10^5$ && 10 & $10240 \times  1024 $&& 0.53\\
$1\times10^4$ && $1\times10^{-1}$ && $10$ && $1\times10^5$ && 10 & $10240 \times  1024 $&& 0.49\\
$1\times10^4$ && $5\times10^{-2}$ && $10$ && $5\times10^4$ && 5 & $10240 \times  2048 $&& 0.40\\ \\ 
$1\times10^4$ && $1\times10^{-1}$ && $1$ && $1\times10^5$ && 10 & $10240 \times  1024 $&& 0.46\\
$1\times10^4$ && $1\times10^{-1}$ && $2$ && $1\times10^5$ && 10 & $10240 \times  1024 $&& 0.44\\
$1\times10^4$ && $1\times10^{-1}$ && $5$ && $1\times10^5$ && 10 & $10240 \times  1024 $&& 0.44\\
$1\times10^4$ && $1\times10^{-1}$ && $10$ && $1\times10^5$ && 10 & $ 10240 \times  1024 $&& 0.49\\  
$1\times10^4$ && $1\times10^{-1}$ && $20$ && $1\times10^5$ && 10 & $ 10240 \times  1024 $&& 0.48\\      
 \end{tabular}
 \caption{\label{tab:resdisp} 
 Summary of the parameters employed in the simulations. The governing parameters of the flow (Rayleigh number $\ra$, domain width $L$ and domain aspect ratio $L/\ra$) and the dispersion parameters ($\Delta$ and $r$, see~\eqref{eq:params}) are indicated, as well as the grid resolution employed.
 Finally, the dimensionless growth rate of the mixing layer, $\gamma$, defined in~\eqref{eq:proffit}, is reported for the simulations with $\ra=10^4$. 
}
 \end{table}

\section{Global budgets and mixing indicators}\label{sec:valid}
Exact global conservation equations relative to the transport of the scalar quantity can be derived and employed to investigate the mixing process. 
We consider here the case of an incompressible and dispersive flow governed by equations~\eqref{eq:equ1bis1}-\eqref{eq:disp01ad}, and we derive the first and second order global budgets in \S\ref{sec:budgetfirst} and \S\ref{sec:budgetsecond}, respectively.
Finally, we introduce the degree of mixing, representative of the current mixing state of the system in \S\ref{sec:degmixing}.

\subsection{First order global budget}\label{sec:budgetfirst}
We apply the volume average operator $\langle\cdot\rangle=1/V\int_V\cdot\textrm{ d}V$ to~\eqref{eq:equ1bis1}. 
Using the divergence theorem, the incompressibility of the flow and the no-penetration/periodic boundary conditions, we derive the global flux of solute in the domain: 
\begin{eqnarray}
    \frac{d \langle C\rangle}{d t} &=& 
    \frac{1}{V}\int_{S=\partial V}
    \left[
    \left(\mathsfbi{D}\nabla C \right)_{z=-\ra/2}
    +\left(\mathsfbi{D}\nabla C \right)_{z=+\ra/2}
    \right]\cdot\mathbf{n}\textrm{ d}S \\
    &=& \frac{1}{\ra S}\int_{S}
    \left[
    -\left(D_{zz}\frac{\partial C}{\partial z} \right)_{z=-\ra/2}
    +\left(D_{zz}\frac{\partial C}{\partial z} \right)_{z=+\ra/2}
    \right]\textrm{ d}S \\
    &=& \frac{1}{\ra}\Bigl[F\left(z=\ra/2\right)-F\left(z=-\ra/2\right)\Bigr] , 
    \label{eq:balf01}
\end{eqnarray}
where $D_{zz} = 1+\Delta^{-1}\left[\lvert\mathbf{u}\rvert+(r-1)w^{2}/\lvert\mathbf{u}\rvert\right]$ (see Appendix~\ref{sec:appB}) and $F(z_i)$ is the dispersive flux at the walls, defined as 
\begin{equation}
F(z_i) = \frac{1}{S}\int_S\left[\left(\frac{\partial C}{\partial z}\right)_{z=z_i} +\left(\frac{|\mathbf{u}|}{\Delta}\frac{\partial C}{\partial z}\right)_{z=z_i}\right]\textrm{d}S
    \label{eq:balf02}
\end{equation}
with $S=L_x$ or $S=L_x\times L_y$ in two- and three-dimensions, respectively.
Note that~\eqref{eq:balf02} corresponds to the definition of equation~(10) proposed by \citet{wen2018rayleigh}.
In the present configuration, due to the no-flux condition at the boundaries ($\partial_z C=0$), we have that $F=0$ and then:
\begin{equation}
    \frac{d \langle C\rangle}{d t} = 0 .
\end{equation}

\subsection{Second order global budget}\label{sec:budgetsecond}
Following previous works on convection in semi-infinite layers \citep{hidalgo2012scaling}, Rayleigh-B\'enard flows \citep{otero2004high,hassanzadeh2014wall,depaoli2024heat,hu2024double} and Hele-Shaw flows \citep{letelier2019perturbative,ulloa2022energetics,ulloa2025convection}, we multiply equation~\eqref{eq:equ1bis1} by $C$ and apply the volume average operator $\langle\cdot\rangle$. 
We use the divergence theorem and the incompressibility of the flow to derive the global budget \citep{depaoli2023review}: 
\begin{align}
    \frac{1}{2}\frac{d \langle C^2\rangle}{d t} &= \frac{1}{V}\int_S
    \left[
    \left(C\mathsfbi{D}\nabla C \right)_{z=-\ra/2}
    +\left(C\mathsfbi{D}\nabla C \right)_{z=+\ra/2}
    \right]\cdot\mathbf{n}\textrm{ d}S-
    \langle (\nabla C)\cdot (\mathsfbi{D}\nabla C)\rangle,
    \label{eq:bal01}\\
    &=\frac{1}{\ra}\Bigl[-(CF)_{z=-\ra/2}+(CF)_{z=+\ra/2}\Bigr]-\langle (\nabla C)\cdot (\mathsfbi{D}\nabla C)\rangle
\end{align}
with $\mathbf{n}$ unit vector normal to the boundary.
Since we refer to a Rayleigh-Taylor configuration, i.e., $\partial_z C(z=\pm\ra/2)=0$, we have that 
\begin{equation}
    \frac{1}{2}\frac{d \langle C^2\rangle}{d t} 
    = -\langle (\nabla C)\cdot (\mathsfbi{D}\nabla C)\rangle .
    \label{eq:budg2a}
\end{equation}
Inspired by the GL theory \citep{grossmann2000scaling,grossmann2001thermal,lohse2024ultimate} where the key idea is to spatially split the viscous dissipation rate into boundary layer and bulk contributions, here we will divide the scalar dissipation into two parts, one ascribed to the molecular component and the other one to the dispersive component of the dispersion tensor. 
Expressed in terms of mean scalar dissipation, \eqref{eq:budg2a} reads:
\begin{equation}
    \frac{1}{2}\frac{d \langle C^2\rangle}{d t} 
    =-\frac{1}{\ra}\chi\label{eq:budg2},   
\end{equation}
where 
\refstepcounter{equation}
$$
    \chi = \chi_m + \chi_d    
    ,\quad    
    \chi_m=\ra\langle|\nabla C|^2\rangle
    ,\quad
    \chi_d=\ra\langle (\nabla C)\cdot (\mathsfbi{D}\nabla C)-|\nabla C|^2\rangle .  
    \eqno{(\theequation{\mathit{a},\mathit{b},\mathit{c}})}\label{eq:defdiss}
$$
The components of the mean scalar dissipation introduced in~\eqref{eq:defdiss} are defined as the total, molecular and dispersive contributions, respectively.
Note that in absence of dispersion ($\Delta\to\infty$) we obtain $\mathsfbi{D}=D_m^*\mathsfbi{I}$ and then $\chi_d=0$.
The choice to introduce $\ra$ in the definition of mean dissipation is required to obtain results that are self-similar when different $\ra$ are considered, i.e., to make the results independent of the domain height until the fingers reach the horizontal boundaries \citep{depaoli2019universal}.
In dimensional terms, the components of the mean dissipation defined in~\eqref{eq:defdiss} read:
\begin{align}
    \chi^* &= \chi^*_m + \chi^*_d \quad , \quad\chi^*_m = D_m^*\langle|\nabla^* C^*|^2\rangle =\frac{\mathcal{U}^*(\Delta C^*)^2}{\phi H^*}\chi_m \label{eq:disdimn23de}\\
    \chi^*_d &= \langle (\nabla^* C^*)\cdot (\mathsfbi{D}^*\nabla^* C^*)-|\nabla^* C^*|^2\rangle = \frac{\mathcal{U}^*(\Delta C^*)^2}{\phi H^*}\chi_d.
    \label{eq:defdissdim}    
\end{align}
Note that $\chi_m^*$ in \eqref{eq:disdimn23de} matches to the definition of dissipation used by \cite{depaoli2024towards}.

\subsection{Quantification of mixing}\label{sec:degmixing}
Following the approach proposed by \citet{jha2011quantifying}, we introduce the degree of mixing, a quantity that varies between 0 and 1, and is representative of the current mixing state of the flow. 
The initial state of the system is characterized by two uniform layers having different concentration and divided by a sharp interface, i.e., $C(x,z>0,t=0)=1$ and $C(x,z<0,t=0)=0$.
Ultimately, the system will achieve a well-mixed condition and the concentration field will be uniform, corresponding to $C(x,z,t\to\infty)=1/2$. 
As a results, the mean concentration variance, defined as $\sigma^2=\langle C^2 \rangle - \langle C \rangle^2$, is initially maximum and equal to $\sigma^2(t=0)=\sigma^2_\text{max}=1/4$, and ultimately minimum and corresponding to $\sigma^2(t\to\infty)=0$. 
The variance is representative of the mixing state, which we quantify using the degree of mixing:
\begin{equation}
    M(t) = 1-\frac{\sigma^2(t)}{\sigma^2_\text{max}}.
    \label{eq:m}
\end{equation}
Such definition gives $M(t=0)=0$ when the two layers are perfectly segregated, and $M(t\to\infty)=1$ when a perfect mixing is achieved.
Also in this case, we split the degree of mixing into two contributions, resulting from the partition of the dissipation \citep{grossmann2000scaling,grossmann2001thermal}.
Given the definition of $\sigma^2$ and the budget~\eqref{eq:budg2}, one can rewrite~\eqref{eq:m} as
\begin{align}
    M(t) &= 1-\frac{\sigma^2(t)}{\sigma^2_\text{max}}\nonumber\\
    &=1-\frac{  \langle C^2(0) \rangle-\langle C(t) \rangle^2 + \int_{\langle C^2(0)\rangle}^{\langle C^2(t)\rangle} \text{ d} \langle C^2\rangle}{\sigma^2_\text{max}}\nonumber\\
    &=\frac{2}{\sigma^2_\text{max}\ra}\int_0^t\left(\chi_m + \chi_d\right) \text{ d}\tau = M_m(t) + M_d(t).
    \label{eq:m2}
\end{align}
where the degree of mixing has been split into the molecular and dispersive components, respectively:
\refstepcounter{equation}
$$
    M_m(t) = \frac{2}{\sigma^2_\text{max}\ra}\int_0^t \chi_m  \text{ d}\tau ,\quad M_d(t)=\frac{2}{\sigma^2_\text{max}\ra}\int_0^t \chi_d  \text{ d}\tau.
    \eqno{(\theequation{\mathit{a},\mathit{b}})}\label{eq:m3}
$$

\section{Flow evolution without dispersion ($\Delta \to \infty$)}\label{sec:resndisp}

\begin{figure}
    \centering
    \includegraphics[width=0.92\columnwidth]{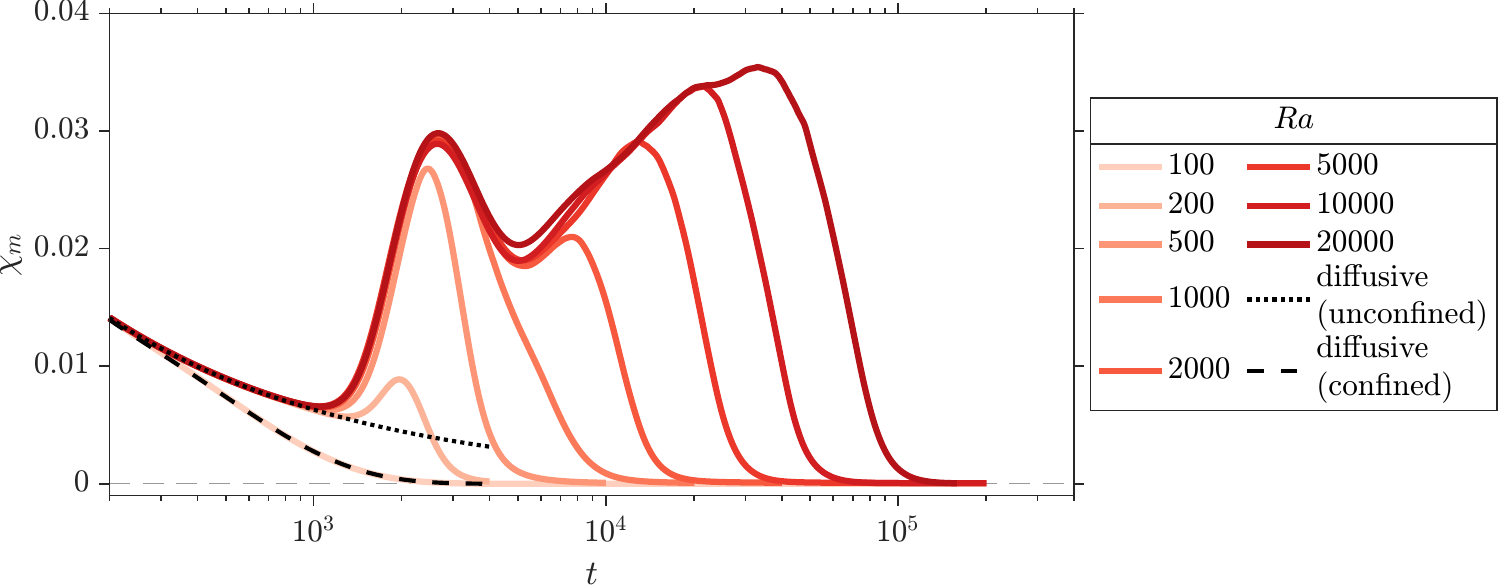} 
    \caption{\label{fig:chira} Evolution of the molecular mean scalar dissipation $\chi_m$ in absence of dispersion and for different Rayleigh numbers, $\ra$ (solid lines). 
    The flow evolution is independent of $\ra$ until flow field of the fingers is significantly influenced by the presence of the horizontal walls.
    The analytical diffusive solutions in the unconfined~\eqref{eq:diff1} (dotted line) and confined~\eqref{eq:ref39appcd} (dashed line) cases are also reported.
    }    
\end{figure}

We consider the simulations performed without dispersion corresponding to $\Delta\to\infty$ in~\eqref{eq:disp01ad}, and we will explore the dynamics for different values of Rayleigh number $\ra$.
We cover a wide range of $\ra$, namely $10^2\le \ra \le 2\times 10^4$ and we simulate domains having constant width $L=10^5$.
The details of all simulation performed are listed in table~\ref{tab:resdisp}.
In absence of dispersion, the influence of $\ra$ on the flow dynamics has been previously investigated in several two-dimensional works \citep{dewit2004miscible,depaoli2019universal,depaoli2019prf,borgnino2021dimensional}.
Here we complement these analyses, which are essential to derive a clear picture of the flow dynamics also in case of dispersive flows.
For all $\ra$ considered, we report in figure~\ref{fig:chira} the evolution of the molecular component of the mean scalar dissipation $\chi_m$ (for $\Delta\to\infty$, $\chi_d=0$).
The dimensionless set of variables used is particularly suitable to highlight the self-similar behaviour of the system.
In particular, the flow evolution is independent of $\ra$ until the domain walls have an effect on the flow, i.e., until the fingers grow and their flow field feels the influence of the horizontal impermeable walls. 
For very small values of $\ra$ ($\ra\le10^3$), the dynamics is soon influenced by the presence of the walls, and convective instabilities may not form or only partially develop.

\begin{figure}
    \centering
    \includegraphics[width=0.99\columnwidth]{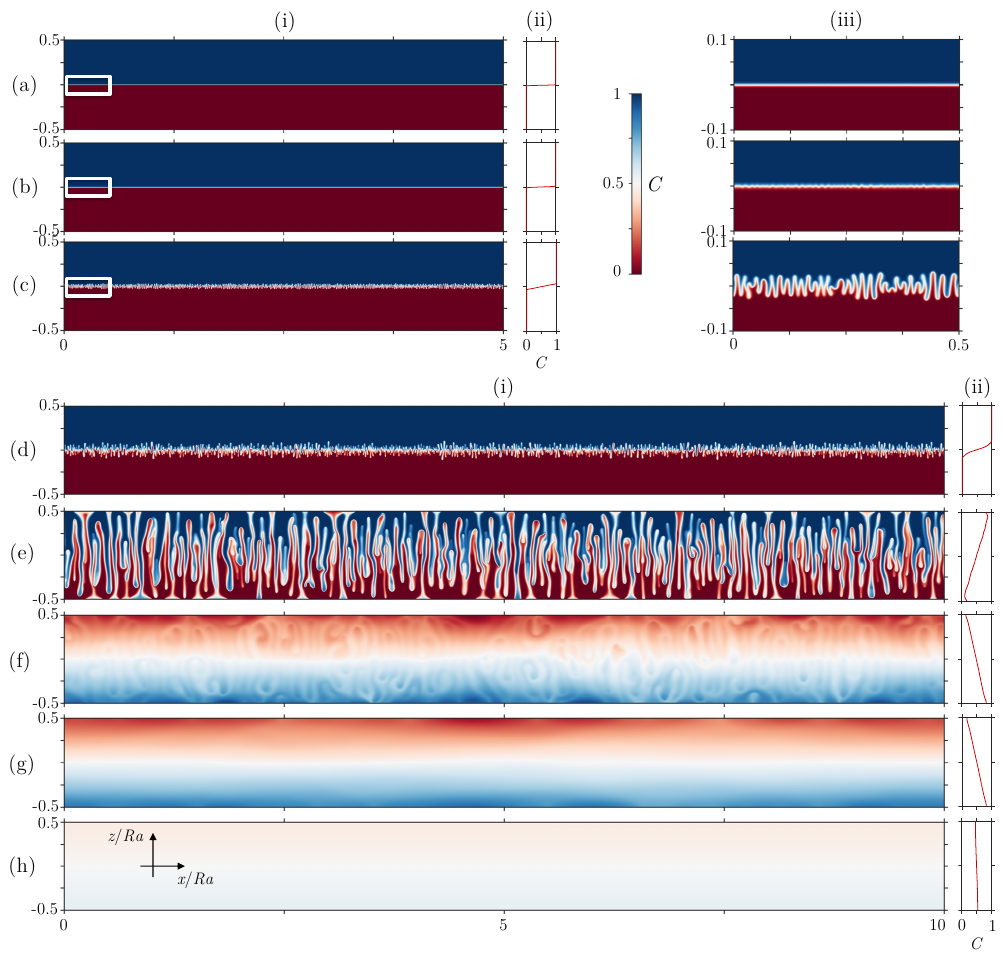}     
    \caption{
    Evolution of the concentration field relative to the simulation $\ra=10^4$. 
    (a-e)~A portion of the domain is shown, corresponding to half of the domain width (left panels, indicated with i) and 1/20 of the domain width (right panels, indicated with ii and corresponding to the white rectangle in the corresponding panels i). 
    The entire domain simulated is shown in panels (d-h). 
    The data correspond to the points indicated in figure~\ref{fig:darcynd}. 
    }
    \label{fig:fields1}
\end{figure}

\begin{figure}
    \centering
    \includegraphics[width=0.80\columnwidth]{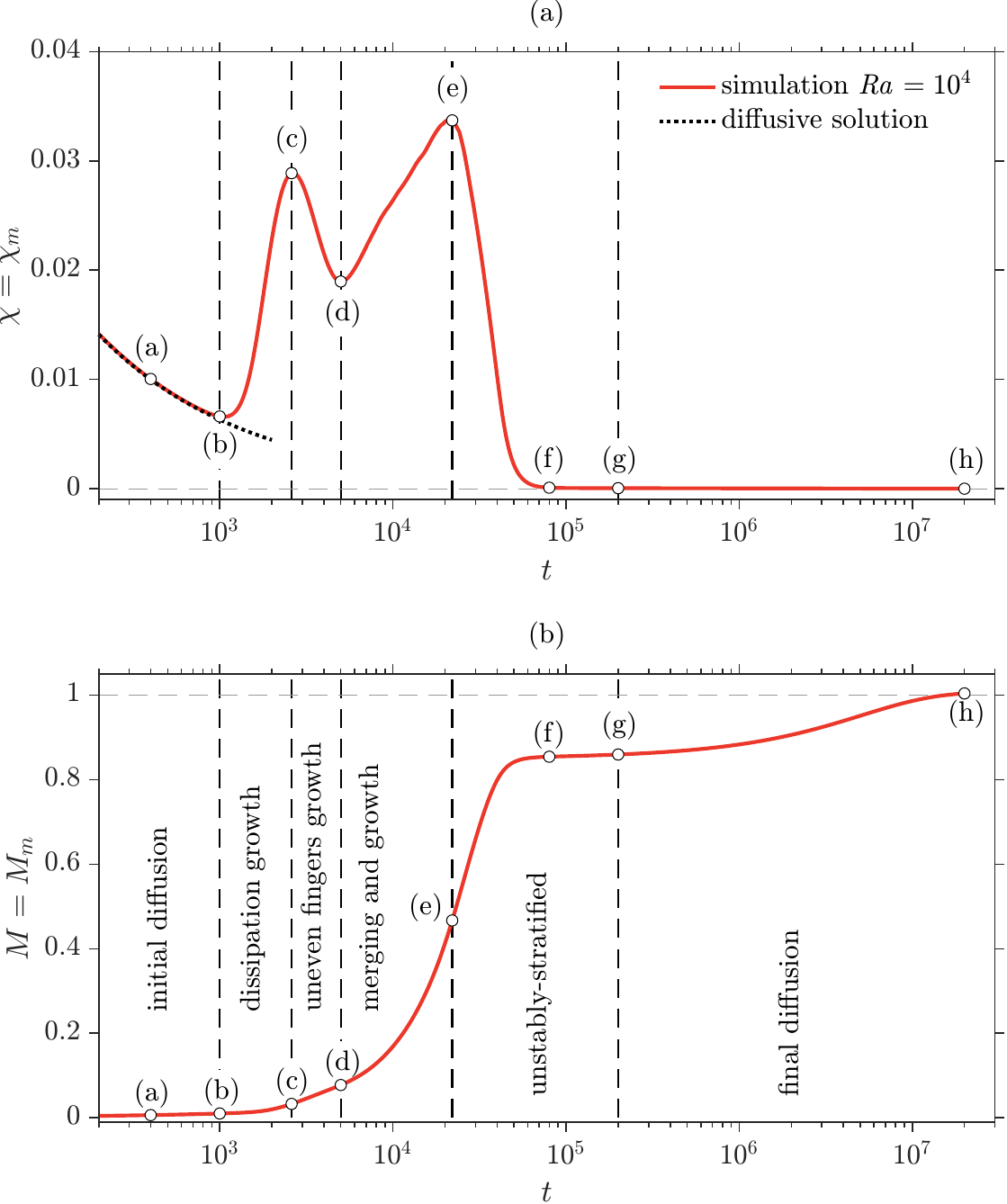}   
    \caption{Evolution of the mean scalar dissipation for the simulation $\ra=10^4$ without dispersions (see table~\ref{tab:resdisp} for additional details).
    The concentration fields and profiles in correspondence of the instants indicated by the letters is shown in figure~\ref{fig:fields1}.
    The diffusive solution~\eqref{eq:diff1} is also reported in~(a).
    The flow regimes identified 
    are indicated in~(b).
    }
    \label{fig:darcynd}
\end{figure}

In the following, to analyse in detail the flow dynamics, we will focus on the simulation at $\ra=10^4$, which serves also as a reference for the dispersive cases studied in \S\ref{sec:delta} and \S\ref{sec:r}.
The system behaviour relative to different $\ra$ can be inferred from the analysis presented.
The flow evolution is initially characterized by inspecting the concentration fields and the concentration profiles in figure~\ref{fig:fields1}, together with the corresponding mixing indicators in figure~\ref{fig:darcynd}.
The process is also illustrated in Movie~\textcolor{red}{S1} of the electronic supplementary material.
We identified several flow regimes characterizing the evolution of the system, which we discuss individually in the following.

\begin{figure}
    \centering
    \includegraphics[width=0.99\linewidth]{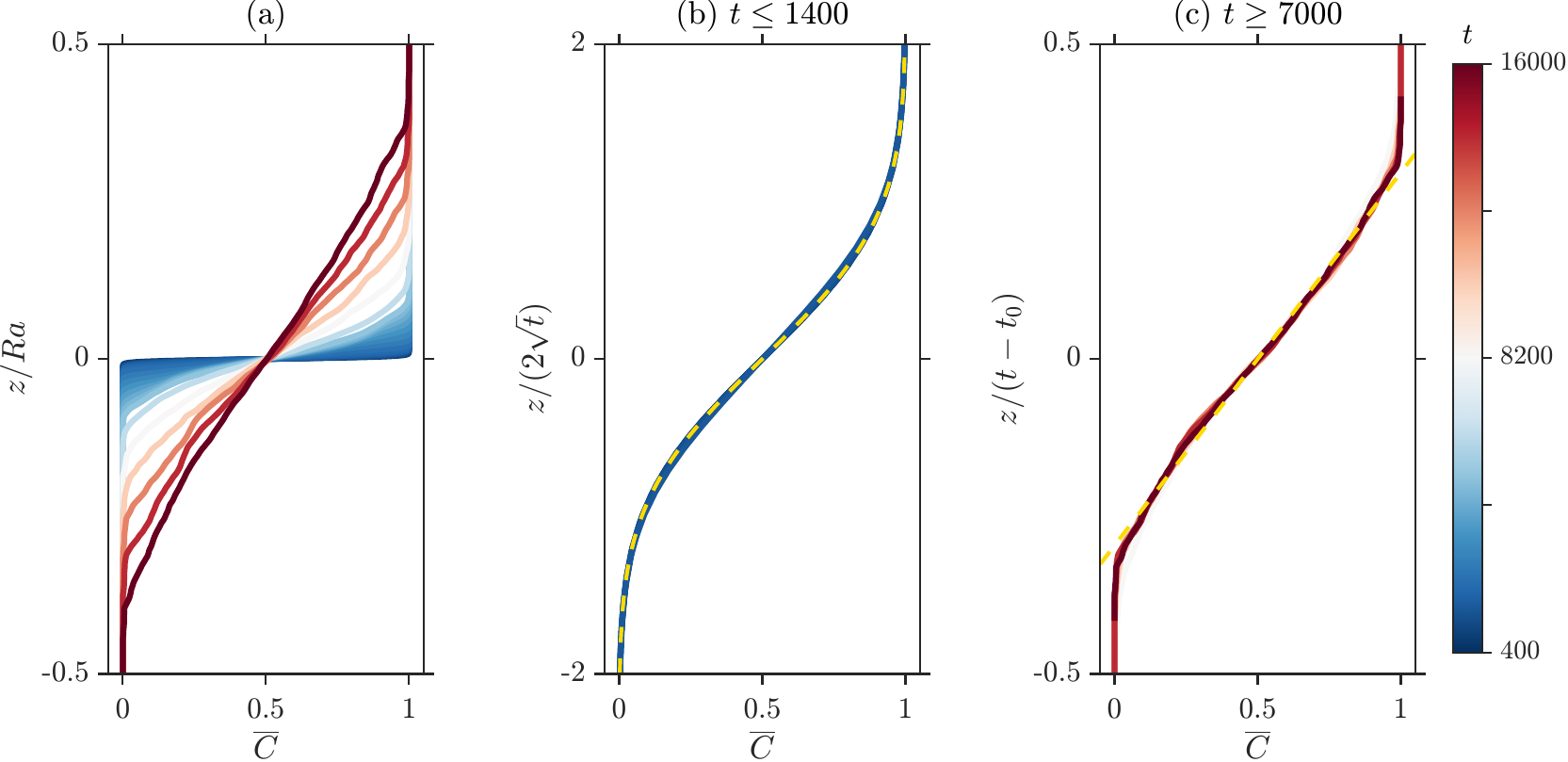} 
    \caption{Evolution of the horizontally-averaged concentration profiles, $\overline{C}$, relative to simulation $\ra=10^4$, $\Delta\to\infty$.
    Profiles reported correspond to instants taken preceding the fingers impact on the walls.
    Specifically, they are in the range $400\le t\le16000$ in (a), in the initial diffusive regime ($t\le1.4\times 10^3$)~in (b) and in the fingers merging and growth regime ($t\ge7\times10^3$) in (c). 
    In panel~(b), the dashed line indicates the initial diffusive solution~\eqref{eq:diffsol}.
    In panel~(c), the wall-normal coordinate is rescaled with $t-t_0$, where $t_0=4\times10^3$. 
    The dashed line represents~\eqref{eq:proffit}.
    }
    \label{fig:profiles_nodisp}
\end{figure}

The flow is initialized with a step-like concentration profile (figure~\ref{fig:fields1}a) with a thin interface dividing the two fluid layers and no flows ($\mathbf{u}=0$).
Therefore, the interface will initially grow driven by diffusion, making the concentration gradient (and then the mean scalar dissipation $\chi=\chi_m$, see figure~\ref{fig:darcynd}a) progressively reduce.
This trend continues until the first convective instabilities appear at $t\approx10^3$, see figure~\ref{fig:fields1}(b-iii).
Similarly to what has been previously observed in convection in semi-infinite domains \citep{elenius2012time}, the precise time at which this growth starts, as well as the maximum values of scalar dissipation later achieved, depend on the initial perturbation. 
However, the overall dynamics is qualitatively comparable to the one described here for the initial condition considered discussed in detail in Appendix~\ref{sec:appA2}.
The flow is initially purely controlled by diffusion.
An analytical solution of the advection-dispersion equation~\eqref{eq:equ1bis1} can be obtained for an unconfined domain, in absence of convection ($\mathbf{u}=0$) and assuming homogeneity in wall-parallel directions ($\partial_x \cdot = 0$), which reads \citep{depaoli2019prf}: 
\begin{equation}
C(z,t)=\frac{1}{2}\left[1+\text{erf}\left(\frac{z}{2\sqrt{t}}\right)\right].
\label{eq:diffsol}
\end{equation}
Results relative to the horizontally-averaged concentration profiles reported in figure~\ref{fig:profiles_nodisp}(b) indicate an excellent agreement with~\eqref{eq:diffsol} (dotted line). 
Employing~\eqref{eq:diffsol} in the definition~\eqref{eq:defdiss}, it follows that the contributions of the mean scalar dissipation during the initial phase (assuming $\mathbf{u}=0$) are
\refstepcounter{equation}
$$
\chi_m = \frac{1}{\sqrt{8\pi t}}, \quad
\chi_d = 0, \quad 
\chi = \frac{1}{\sqrt{8\pi t}},
\eqno{(\theequation{\mathit{a},\mathit{b},\mathit{c}})}\label{eq:diff1}\\
$$
and they are very well captured by the simulation (see figure~\ref{fig:darcynd}a).
Correspondingly, equation~\eqref{eq:m3} predicts the degree of mixing to evolve as
\refstepcounter{equation}
$$
M_m =\frac{8\sqrt{t}}{\sqrt{2\pi}} , \quad
M_d = 0 , \quad
M = \frac{8\sqrt{t}}{\sqrt{2\pi}}. 
\eqno{(\theequation{\mathit{a},\mathit{b},\mathit{c}})}\label{eq:diff2}
$$
Note that for very low Rayleigh numbers ($\ra=10^2$), the solution~\eqref{eq:diff1} does not represent a good approximation due to the effect of confinement: a no-flux boundary condition should be employed to determine the analytical solution.
In Appendix~\ref{sec:appC}, we derive the analytical diffusive solution~\eqref{eq:ref39appcd} relative to the confined case, which we reported in figure~\ref{fig:chira} (dashed line), which describes very well the evolution observed numerically.

Similarly to what has been observed in semi-infinite domains \citep[see][and references therein]{slim2014solutal,depaoli2025grl}, the flow is later characterized by a dissipation growth (figures~\ref{fig:fields1}b-c):
When the newly-formed fingers have grown, they accelerate vertically sharpening the concentration gradient at their interface, corresponding to the increase in the mean dissipation observed in figure~\ref{fig:darcynd}(a) between $t\approx10^3$ and $t\approx2.5\times10^3$.

Later, fingers grow further, but in an uneven manner (figures~\ref{fig:fields1}c-d): some fingers extend vertically at large velocity and invade regions of the domain with uniform concentration, and far from other fingers. 
In this situation, the strong concentration difference between the inner and outer fingers region, combined to the absence of neighbouring flow structures (and then no strong shear) prevents the fingers from expanding laterally, and makes the interface to grow diffusively while the fingers continue to extend vertically. 
As a result, the concentration gradients reduce and so does and the mean scalar dissipation in figure~\ref{fig:darcynd}(a).

The velocity field generated by each finger induces interactions with the neighbouring ones, eventually leading to merging:
The pioneering fingers perturb the velocity field of the neighbouring shorter fingers, compressing and forcing them to retreat and eventually to merge  (figure~\ref{fig:fields1}d-e).
In this phase, similarly to what has been discussed in the uneven fingers growth regime, the fingers keep growing vertically in unexplored regions of the flow, but now there is a large and nearly uniform reservoir of solute at the core of the fingers: The gradient of concentration at the fingers interface still decreases, but at a lower rate compared to the increase of interfacial extension of the fingers, driven by their vertical growth.
As a result, the overall behaviour corresponds to an increase of the mean scalar dissipation in figure~\ref{fig:darcynd}(a).

During the fingers merging and growth phase, the mixing region grows approximately linearly in time \citep[namely, $\sim t^{1.2}$, see][]{depaoli2019prf,depaoli2022experimental}.
Indeed, rescaling the profiles figure~\ref{fig:profiles_nodisp}(c) with respect to $z/t$, they result to be well approximated by the linear function 
\begin{equation}
    \overline{C} = \frac{1}{2}+\frac{z}{\gamma(t-t_0)}
    \label{eq:proffit}
\end{equation}
(dashed line in figure~\ref{fig:profiles_nodisp}c), where $t_0=4\times10^3$ corresponds (approximately) to the time at which this regimes starts, and $\gamma=0.59$ is the dimensionless growth rate of the mixing layer, in good agreement with the measurements of \citet{boffetta2022dimensional}, who reported a value of 0.67 obtained at larger $\ra$.
The value of $\gamma$ computed for all the simulations is reported in table~\ref{tab:resdisp}.
It has been obtained as a best fit of~\eqref{eq:proffit} for the concentration profile within the range $0.05\le C\le 0.95$, with $t_0=4\times10^3$ and within the interval $7\times10^3 \le t \le 1.6\times 10^4$.
For simulations $\Delta=5\times10^{-2}$ and $r=20$, in which the fingers development is slower, the profiles are considered in the interval $1.5\times10^4 \le t \le 2.7\times 10^4$.

When the fingers reach the horizontal boundaries of the domain ($t\approx2\times10^4$), the growth of mean scalar dissipation arrests.
At this time, the concentration profile is nearly linear and still unstable (figure~\ref{fig:fields1}e-ii), indicating that the upper portion of the system is still characterized by a higher concentration of solute, and then larger density, compared to the lower layer.
The density contrast between the upper and lower domain, however, is much smaller than at the beginning. 
There is no further fresh fluid available to mix with the fluid at the core of the fingers, and the local concentration gradient across the interface of the fingers decreases progressively.
At the same time, the situation at the domain centre ($z=0$) is steady, with no new fingers forming and no fingers merging, and with the solute being transported along the fingers path already created.
This process occurs from the unstably-stratified configuration in figure~\ref{fig:fields1}(e) until a stably-stratified solute distribution is achieved (figure~\ref{fig:fields1}g).
Note that the footprint of the fingers persists for a long time (figure~\ref{fig:fields1}f), and it smoothly disappears when the residual convective driving dies out. 
Indeed, despite being separated by approximately $1.2\times10^5$ dimensionless time units, the difference between the horizontally-averaged concentration profiles in figure~\ref{fig:fields1}(f) and figure~\ref{fig:fields1}(g) is very limited.

\begin{figure}
    \centering
    \includegraphics[height=0.47\columnwidth]{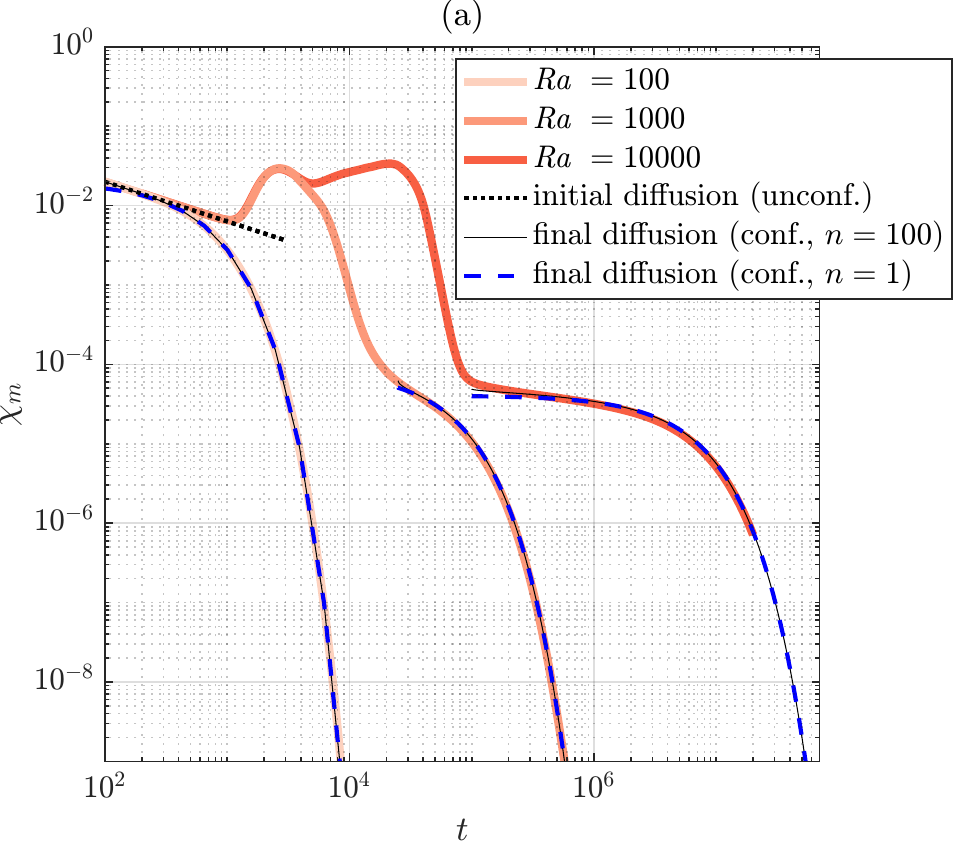}
    \includegraphics[height=0.47\columnwidth]{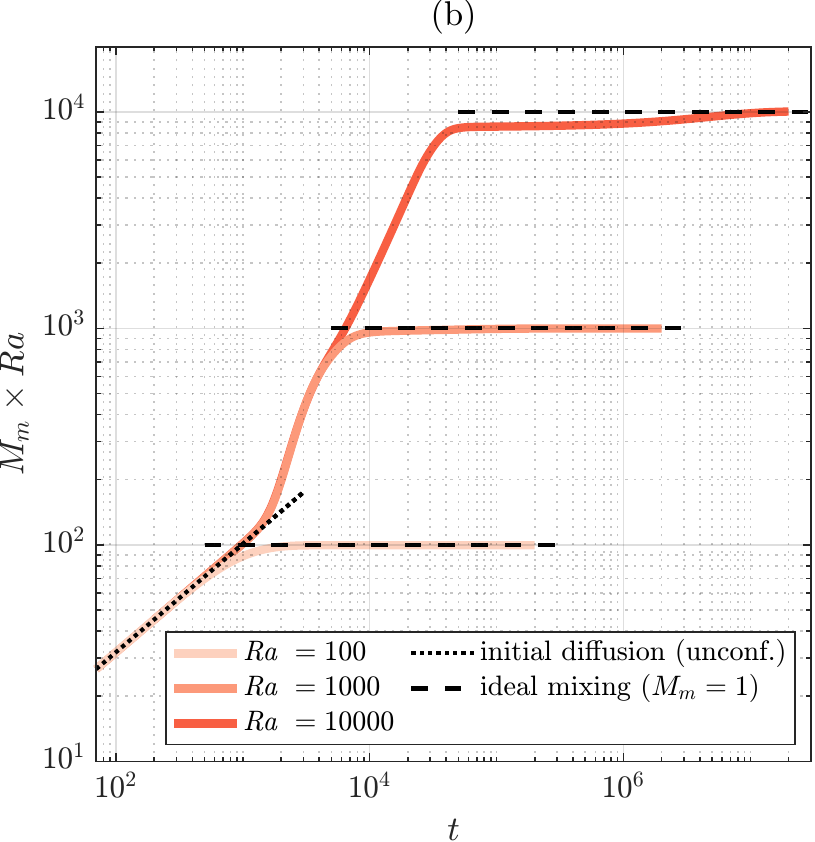}
    \caption{
    Scalar dissipation rate (a) and rescaled degree of mixing~(b) are reported as a function of time $t$ for three values of Rayleigh number, $\ra$, namely $10^2, 10^3, 10^4$. 
    The system is self-similar and at early times it follows the analytical solutions (initial unconfined diffusion), \eqref{eq:diff1} for $\chi_m$ and \eqref{eq:diff2} for $M_m$, indicated here with black dotted lines. 
    As soon as the system achieves a stably-stratified condition, it enters the final diffusive phase.
    The scalar dissipation evolves according to~\eqref{eq:ref39appcd2} indicated in~(a) by the black solid lines and computed using $n=100$ (note that for $\ra=10^2$, the initial confined solution~\eqref{eq:ref39appcd} is used).
    However, these solutions are very well approximated also when $n=1$, scales as~\eqref{eq:ddr55} and is indicated by the blue dashed lines.
    Ultimately, the domain attains the fully mixed condition~\eqref{eq:ddr6} (black dashed line in panel~b).
    }
    \label{fig:darcydr}
\end{figure}

When the stably-stratified distribution (figure~\ref{fig:fields1}g-ii) is achieved, the buoyancy effects are negligible, and diffusion drives mixing from now on.
The system enters a final diffusive regime that, despite being characterized by a very low value of the mean dissipation (figure~\ref{fig:darcynd}a), it is responsible for approximately 14\% of the overall mixing, i.e. (see figure~\ref{fig:darcynd}b) $M$ grows from $\approx0.86$ in~(g) to $\approx1$ in~(h).
Similarly to what we have derived for $\ra=10^2$ for the initial confined diffusive evolution (see Appendix~\ref{sec:appC}), we can derive here an analytical solution for the mean scalar dissipation during the late stage.
Again, we use~\eqref{eq:equ1bis1}, assume that the fluid is still ($\mathbf{u}=0$) and consider the system uniform in horizontal direction ($\partial_x C = 0$).
At large $\ra$ a different initial condition has to be considered, namely a linear concentration profile, representative of a stably stratified flow that starts at $t=t_f$ (see Appendix~\ref{sec:appC2} for addition details).
The solution \eqref{eq:ref39appcd2} computed using $n=100$ is shown in figure~\ref{fig:darcydr}(a) for 3 values of $\ra$, namely $\ra=10^2$, $10^3$ and $10^4$ (with the case $10^4$ previously discussed throughout the entire flow evolution).
The accuracy of the analytical solution in predicting the decay of dissipation obtained in the simulations is excellent. 
As already observed in the initial confined diffusion for $\ra=10^2$, also in~\eqref{eq:ref39appcd2} we have that $\partial C/\partial \bar{z}\sim\exp(-n^2)$, indicating a fast decay with $n$.
Therefore, in leading order the dissipation can be approximated as
\begin{equation}
\chi_m(t)\sim
\exp{\left[-2\left(\frac{\pi}{\ra}\right)^2t\right]}\\
\label{eq:ddr55}
\end{equation}
which we show in figure~\ref{fig:darcydr}(a) and indicate with $n=1$.
Also in this case, the agreement is excellent, and we conclude that equation~\eqref{eq:ddr55} can be employed to approximate the behaviour during the final diffusive phase.
For long times, the well-mixed condition is achieved, corresponding to the asymptotic equilibrium profile:
\begin{equation}
C(\bar{z},t\to\infty) = 1/2,
\label{eq:ddr3}
\end{equation}
with the mean scalar dissipation~\eqref{eq:ref39appcd2} and degree of mixing equal to: 
\refstepcounter{equation}
$$
\chi_m(t\to\infty)=0\quad\text{ and}\quad M_m(t\to\infty)=1,
\eqno{(\theequation{\mathit{a},\mathit{b}})}\label{eq:ddr6}
$$
respectively.
One can indeed observe in figure~\ref{fig:darcydr}(b) that the well-mixed condition is ultimately achieved.
Note that, for ease of comparison, the degree of mixing $M_m$ is shown rescaled by the $\ra$, and the corresponding diffusive solution is $M_m\times\ra = 8\ra\sqrt{t/(2\pi)}$.

\section{Flow evolution with dispersion: influence of $\Delta$}\label{sec:delta}

\begin{figure}
    \centering
    \includegraphics[width=0.99\linewidth]{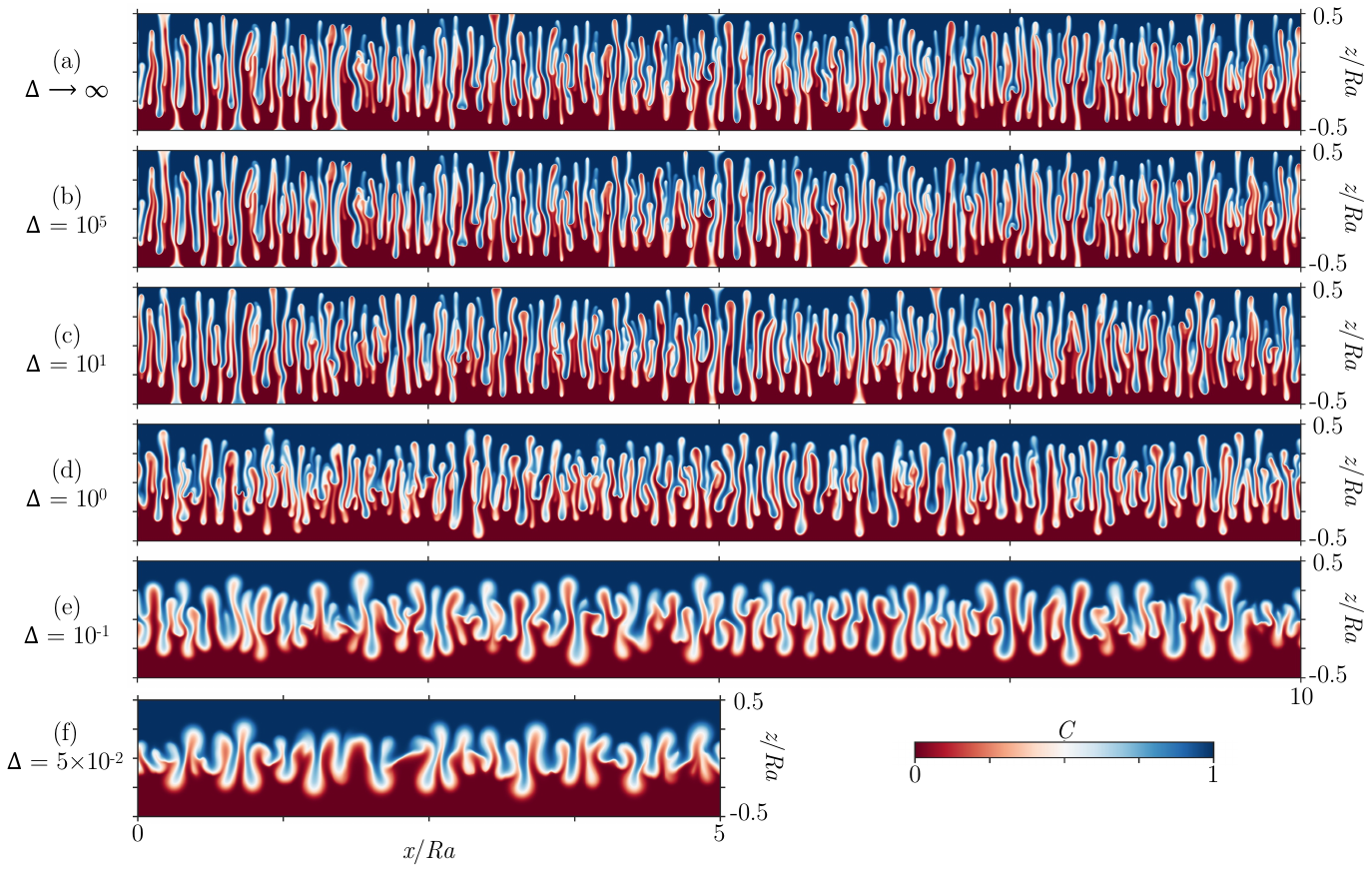}
    \caption{Concentration fields at $t=2\times10^4$ for different values of the dispersion parameter $\Delta$. The field~(a) corresponds to the case without dispersion ($\Delta\to\infty$).
    See Movie~\textcolor{red}{S2} of the electronic supplementary material for the time-dependent evolution of the simulation with $\Delta=10^{-1}$ (case e).
    }
    \label{fig:fieldsdisp}
\end{figure}

We present here the results relative to the effect of $\Delta$ at $\ra=10^4$ and $r=10$.
The choice of $\ra$ is motivated by the fact that at such $\ra$ all the phases of the flow evolution described in \S\ref{sec:resndisp} can be observed. 
Since the dispersion model~\eqref{eq:disp01ad} considered in this work relies on two parameters, $\Delta$ and $r$ defined in \eqref{eq:params}, to keep the computational costs affordable, we decided to vary $\Delta$ and fix the anisotropy ratio to $r=10$.
Determining which value is more appropriate for $r$ is a debated matter \citep{delgado2007longitudinal}, since the longitudinal ($D_l^*$) and transverse ($D_t^*$) components of dispersion depend on several flow parameters \citep[e.g., Schmidt number, Reynolds number, tortuosity of the medium, P\'eclet number, fluid phases involved, see][and references therein]{depaoli2023review}.
The value of $r$ varies with the dispersion regime, and in general it is not appropriate to take $D_t^*$ to be one order of magnitude smaller than $D_l^*$, which is the most common choice in numerical studies of dispersion.
However, when studying solute transport in the advection dominated regime, this assumption holds \citep{bijeljic2007pore}, and therefore we consider our findings representative of this case.

\subsection{Flow dynamics}\label{sec:delta_dyn}

In presence of dispersion, the evolution of the flow follows a behaviour that is similar to the case discussed in \S\ref{sec:resndisp}, and eventually with fewer regimes as $\Delta$ is decreased (see Movie~\textcolor{red}{S2} of the electronic supplementary material for the time-dependent evolution of the simulation with $\Delta=10^{-1}$, reported in figure~\ref{fig:fieldsdisp}e). 
The concentration distribution over the entire field at $t=2\times 10^4$ is reported in figure~\ref{fig:fieldsdisp} for different values of $\Delta$.
The effects of $\Delta$ on the flow morphology, which will be discussed in the following, are multiple.

\begin{figure}
    \centering
    \includegraphics[width=0.99\linewidth]{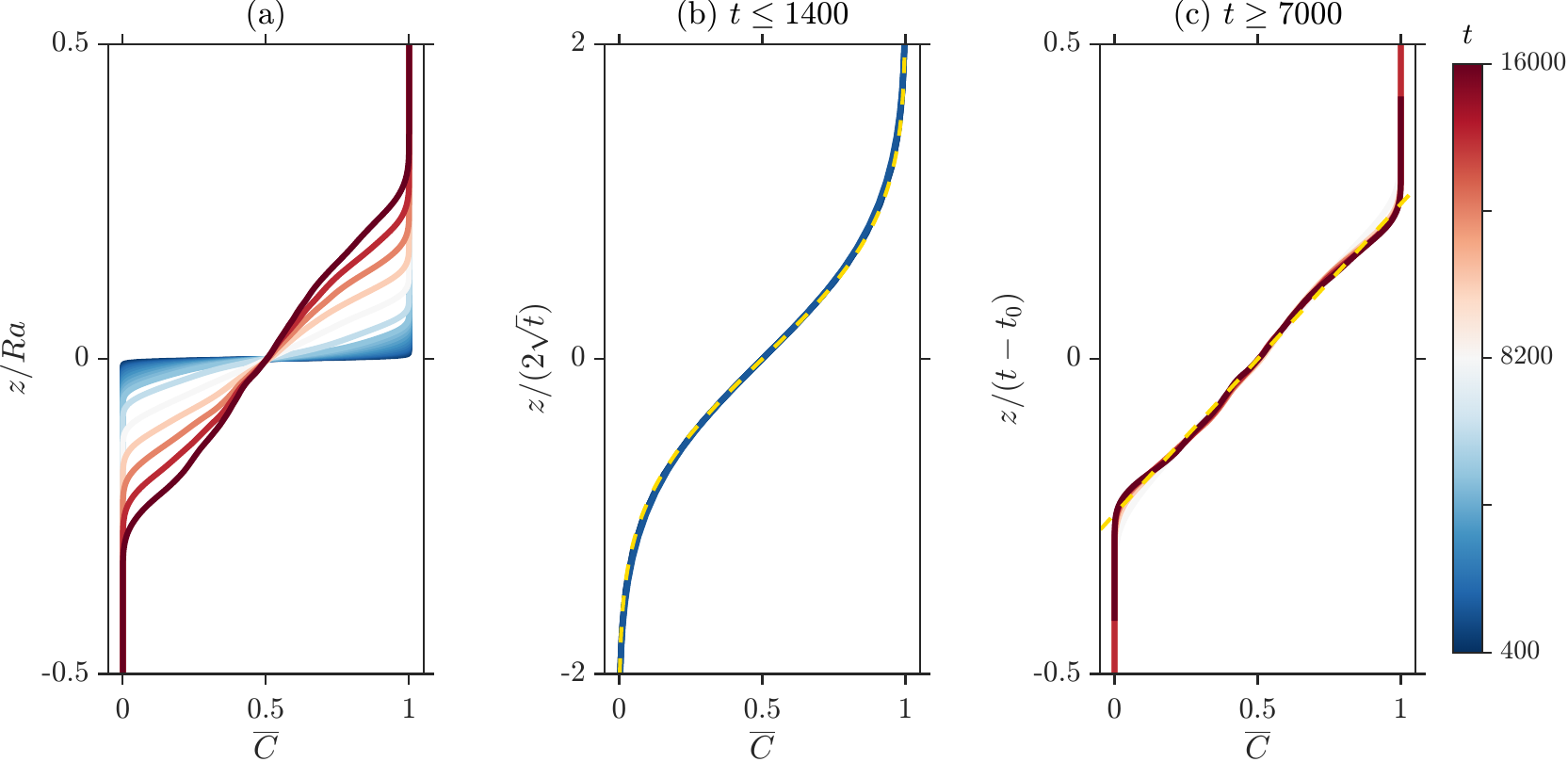} 
    \caption{Evolution of the horizontally-averaged concentration profiles, $\overline{C}$, relative to simulation $\ra=10^4$, $\Delta=0.1$ and $r=10$.
    Profiles reported correspond to instants taken preceding the fingers impact on the walls.
    Specifically, they are in the range $400\le t\le16000$ in (a), in the initial diffusive regime ($t\le1.4\times 10^3$)~in (b) and in the fingers merging and growth regime ($t\ge7\times10^3$) in (c). 
    In panel~(b), the dashed line indicates the initial diffusive solution~\eqref{eq:diffsol}.
    In panel~(c), the wall-normal coordinate is rescaled with $t-t_0$, where $t_0=4\times10^3$. 
    The dashed line represents~\eqref{eq:proffit}.
    }
    \label{fig:profiles_delta}
\end{figure}

Results relative to the horizontally-averaged concentration profiles are reported in figure~\ref{fig:profiles_delta}, where different times, namely $400\le t\le 16000$, are shown for the simulation with $\Delta=0.1$ and $r=10$ (figure~\ref{fig:profiles_delta}a).
At early times, figure~\ref{fig:profiles_delta}(b), the profiles follow the diffusive growth (\eqref{eq:diffsol}, dashed line). 
Later, the mixing region extends approximately linearly in time and the profiles are very well fitted by the linear function~\eqref{eq:proffit} (dashed line in figure~\ref{fig:profiles_delta}c), where $t_0=4\times10^3$ and $\gamma=0.49$.
Note that in this case the dimensionless growth of the mixing layer $\gamma$, is lower compared to the case without dispersion ($\gamma=0.59$).
This can be easily verified comparing the concentration fields of figures~\ref{fig:fieldsdisp}(a) and~\ref{fig:fieldsdisp}(e).

\begin{figure}
    \centering
    \includegraphics[width=0.99\linewidth]{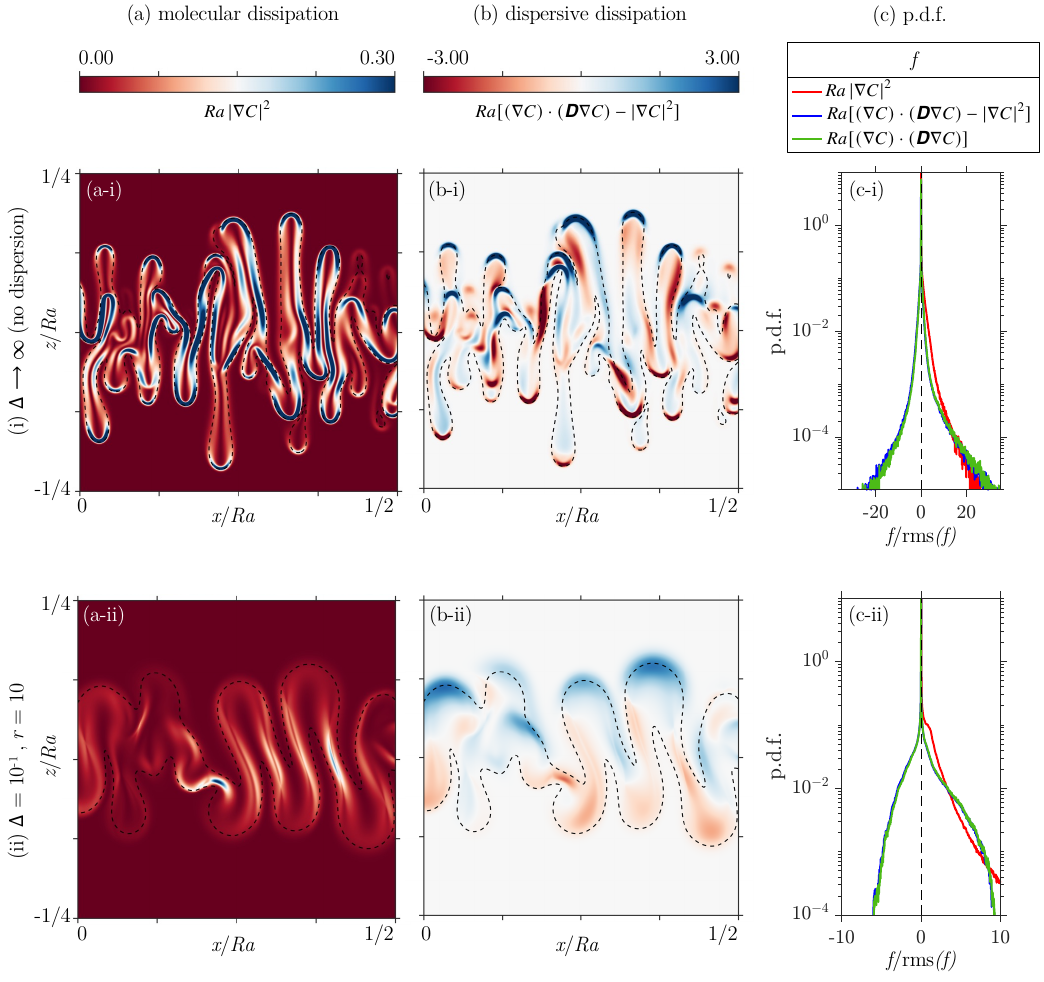}
    \caption{
    The distributions of molecular ($\ra|\nabla C|^2$, see~\eqref{eq:defdiss}) and dispersive ($\ra[ (\nabla C)\cdot (\mathsfbi{D}\nabla C)-|\nabla C|^2]$) scalar dissipation are shown in panels~(a) and (b), respectively, taken at time $t=10^4$. 
    The case without dispersion is shown in (i) ($\ra=10^4,\Delta\to\infty$), and a case with dispersion in (ii)~($\ra=10^4,\Delta=10^{-1},r=10$).
    For better visualization, only a small region in the core of the domain is shown ($0\le x/\ra \le 1/2, -1/4\le z/\ra \le 1/4$).
    The dashed lines represent the iso-contours $C=1/4$ and $C=3/4$.
    The probability density function (p.d.f) of the components of the molecular, dispersive and total dissipation normalized by their respective root mean squares (rms) and relative the specific fields considered are reported in panels~(c).
    For better visualization the limits of the colorbars in (a) and (b) are reduced compared to the maximum/minimum values present in the field.
    }
    \label{fig:fieldpdf}
\end{figure}

A first macroscopic observation is that reducing $\Delta$ has the effect of diminishing the number of fingers.
This is apparent for $\Delta\le1$ (figures~\ref{fig:fieldsdisp}d-f), but it occurs also for $\Delta=10$ (figure~\ref{fig:fieldsdisp}c), while no difference is observed when $\Delta=10^5$ (figure~\ref{fig:fieldsdisp}b) compared to the case without dispersion ($\Delta\to\infty$, figure~\ref{fig:fieldsdisp}a).
Since for small $\Delta$ the neighbouring fingers are further away compared to the case without dispersion, each finger has more space available to diffusively grow horizontally.
This morphological effect combines with the additional spreading provided by dispersion, which further reduces the concentration gradient across the fingers interface.
A visual interpretation of this mechanisms is reported in figure~\ref{fig:fieldpdf}.
The distribution of the molecular component of the scalar dissipation ($\ra|\nabla C|^2$) is shown in figures~\ref{fig:fieldpdf}(a-i) and \ref{fig:fieldpdf}(a-ii), for a case without dispersion ($\ra=10^4,\Delta\to\infty$) and with dispersion ($\ra=10^4,\Delta=10^{-1},r=10$), respectively.
The concentration fields refer to time $t=10^4$, at which the fingers have grown to less than half domain height (only a small region in the core of the domain is shown in figures~\ref{fig:fieldpdf}a-b).
In absence of dispersion ($\Delta\to\infty$, figure~\ref{fig:fieldpdf}a-i), the fingers are constrained to grow vertically within the narrow space between two neighbouring fingers.
This produces large values of dissipation at their side interface and also at their tips. 
Other regions of high dissipation are identified as the portion of domain near the centreline, where the fingers pattern is more complex. 
In the dispersive case ($\Delta=10^{-1}$, figure~\ref{fig:fieldpdf}a-ii), high values of dissipation are still localized in the same regions (side interface, tips and near the domain centreline), but corresponding values are much lower. 
This matter will be further discussed from a global perspective in \S\ref{sec:delta_mix}.

A long term consequence of the lower growth of the mixing region ($\gamma=0.49$) compared to the case without dispersion ($\gamma=0.59$) is that it will take longer for the fingers to reach the walls, and then to achieve the maximum value of mean scalar dissipation.
When this happens, however, the subsequent dynamics is similar to the one described in the unstably-stratified and final diffusive regimes, leading to a progressive homogenization of the concentration distribution.
In the following section, the global mixing dynamics will be presented.

\subsection{Mixing}\label{sec:delta_mix}

\begin{figure}
    \centering
    \includegraphics[height=0.32\columnwidth,clip=true, trim={0 0 3.5cm 0}]{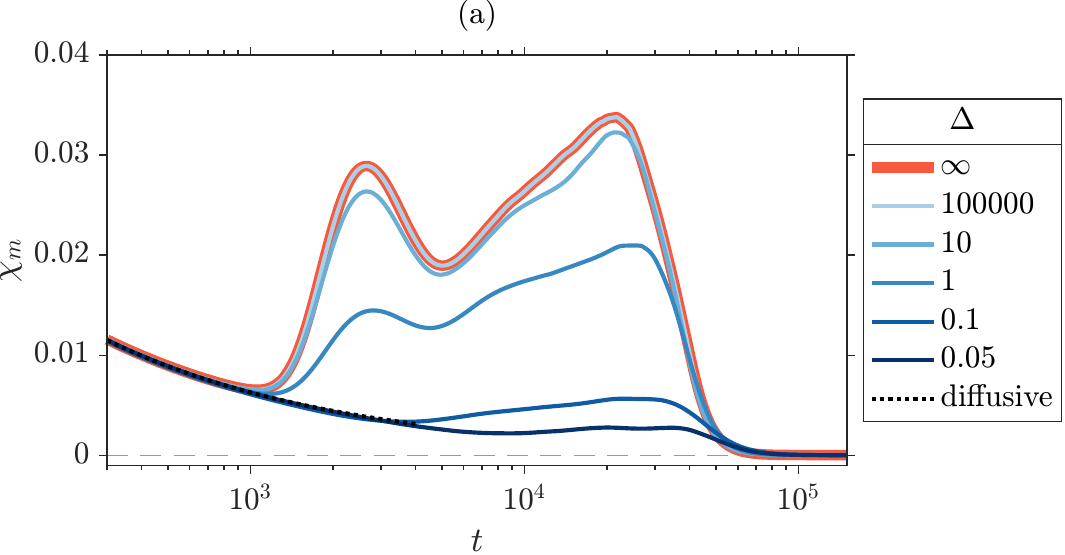}     
    \includegraphics[height=0.32\columnwidth,clip=true, trim={0 0 3.3cm 0}]{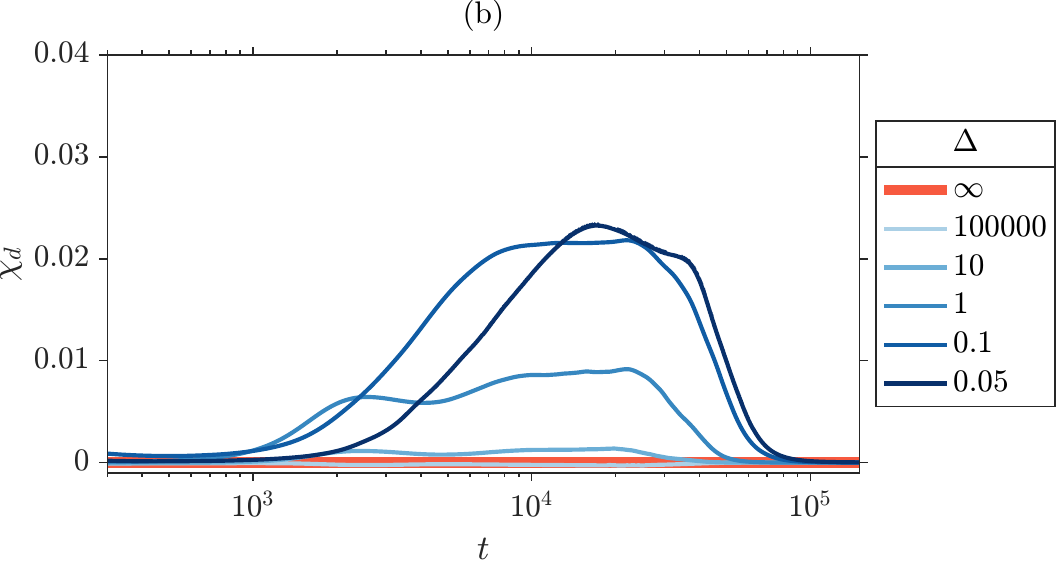}

\vspace{0.5cm}
    
    \includegraphics[height=0.32\columnwidth]{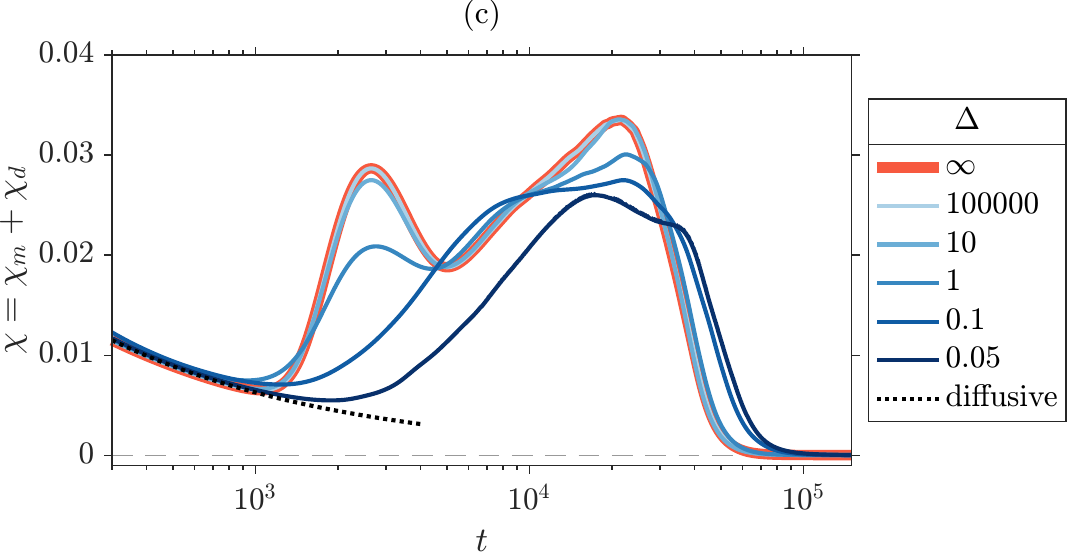}

    \caption{\label{fig:degdelta}  
    Evolution of the mean scalar dissipation for different values of $\Delta$.
    The red line refers to the case in absence of dispersion.
    The molecular ($\chi_m$), dispersive ($\chi_d$) and total dissipation ($\chi=\chi_m+\chi_d$) are reported in panels~(a), (b) and (c), respectively. 
    The initial diffusive solution~\eqref{eq:diff2} (dotted line) is also indicated. 
    }    
\end{figure}

A behaviour similar to that discussed in \S\ref{sec:resndisp} is observed also in this case, with some variations: after an initial diffusion-dominated phase still well described by~\eqref{eq:diff2}, $\chi_m$, reported in figure~\ref{fig:degdelta}(a) decreases when $\Delta$ is reduced, as a result of the thickening of the fingers interface, leading to smoother gradients of concentration.
The same trend was observed in previous studies in different flow configurations \citep{dhar2022convective}.

Understanding the distribution of the dispersive scalar dissipation within the domain, $\ra[ (\nabla C)\cdot (\mathsfbi{D}\nabla C)-|\nabla C|^2]$ reported in figure~\ref{fig:fieldpdf}(b), is less trivial.
In general, large values are found where both concentration gradient and velocity are large.
In absence of dispersion (figure~\ref{fig:fieldpdf}b-i), large (in magnitude, either positive or negative) values of dispersive dissipation correspond to fingers tips and core.
In contrast, when dispersion is considered (figure~\ref{fig:fieldpdf}b-ii), the whole mixing region is characterized by non-negligible values of dispersive dissipation. 
A more quantitative evaluation is provided by the probability density functions (p.d.f.) of the components of the molecular, dispersive and total dissipation relative the specific fields considered, and reported in figure~\ref{fig:fieldpdf}(c).
Globally the results appear very different: without dispersion (figure~\ref{fig:fieldpdf}c-i) the distribution of dispersive dissipation is nearly symmetric and gives zero as mean global value, while in the dispersive case  (figure~\ref{fig:fieldpdf}c-ii) the p.d.f. is skewed towards the positive values.
As a result, we have that the smaller the value of $\Delta$, the larger the values of dispersive dissipation (see figure~\ref{fig:degdelta}b).
Since the flow is initialized as still ($\mathbf{u}=0$), at early times the dispersion tensor~\eqref{eq:disp01ad} corresponds to $\mathsfbi{D}=\mathsfbi{I}$, leading to a dispersive dissipation that is zero.
As soon as convective instabilities form, $\mathsfbi{D}$ grows in magnitude, as well as $\chi_d$.
The convective onset occurs earlier when dispersion is considered (see figure~\ref{fig:degdelta}b and c).
In agreement with previous findings obtained in a different configuration \citep{dhar2022convective}, we find that for this large value of $\ra$ the effect of dispersion on the onset time is not very significant, while it may be much more pronounced at lower $\ra$.
In addition,  the onset time is also non-monotonic with $\Delta$.
Eventually, for long times, the flow achieves a stably stratified configuration, and the strength of convective instabilities diminishes.
As a result, $\mathsfbi{D}$ reduces again leading to a decrease of the total dissipation.

The total mean dissipation $\chi$ is obtained combining the molecular ($\chi_m$, figure~\ref{fig:degdelta}a) and dispersive ($\chi_d$, figure~\ref{fig:degdelta}b) contributions, and it is reported in figure~\ref{fig:degdelta}(c).
Despite an earlier onset observed when dispersion is considered, which makes $\chi$ to depart sooner from the initial diffusive dissipation (\eqref{eq:diff2}, dotted line), simulations with large $\Delta$ initially exhibit larger values of total dissipation. 
For instance, it takes about $t\approx8\times10^3$ for the case $\Delta=0.1$ to achieve values of $\chi$ comparable to those at larger $\Delta$, and even later the maximum total dissipation achieved is smaller compared to cases without dispersion. 
However, the situation changes dramatically after the fingers reach the walls, corresponding to the maximum of $\chi$ in figure~\ref{fig:degdelta}(c), between $t\approx2\times10^4$ and $t\approx5\times10^4$, depending on $\Delta$. 
For $\Delta=0.1$ and $\Delta=0.05$, despite $\chi$ being initially smaller compared to larger $\Delta$, the total dissipation continues to contribute to mixing for a much longer time.
Indeed, after the fingers reach the walls, the velocity reduces partially, but not suddenly. 
This provides a significant advantage compared to the cases without dispersion, especially when it comes to the cumulative amount of solute mixed, quantified by $M$.

\begin{figure}
    \centering   

    \includegraphics[width=0.7\columnwidth]{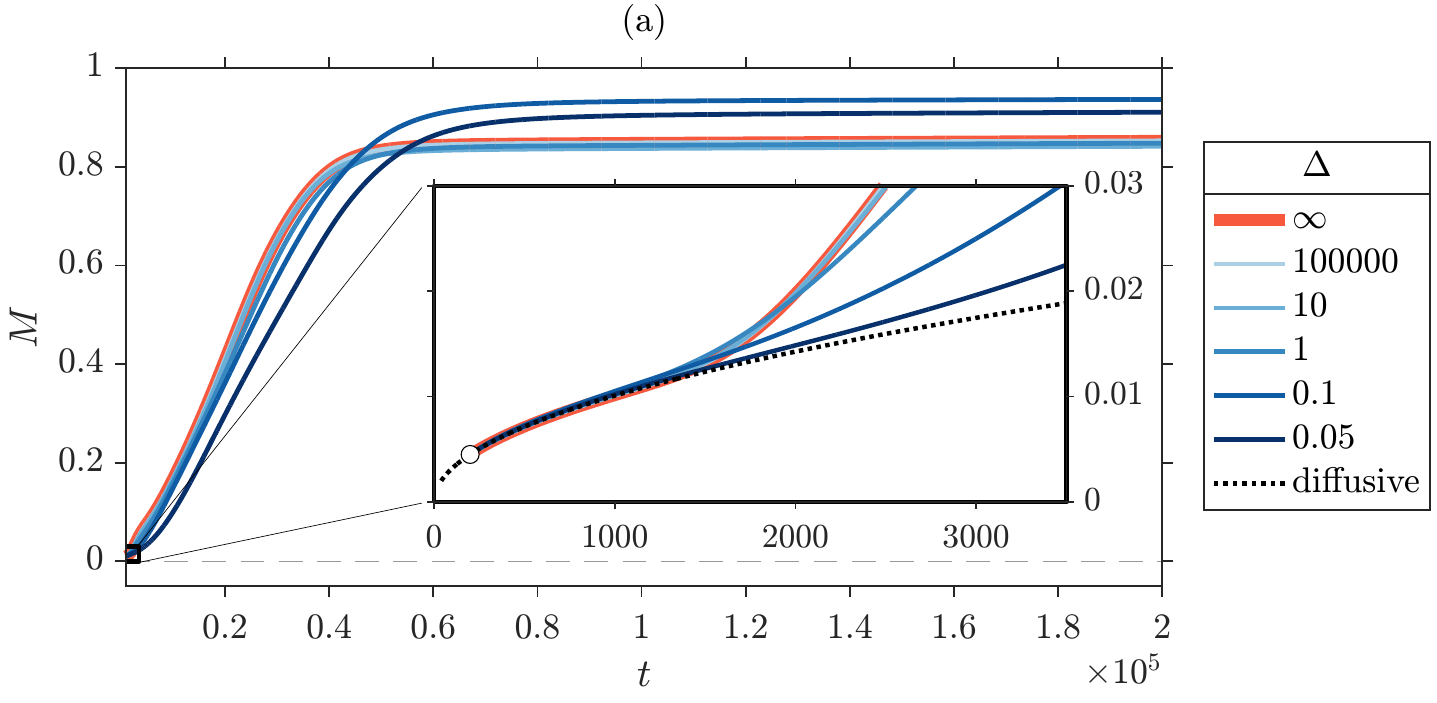}
    
    \vspace{0.5cm}
    
    \includegraphics[width=0.7\columnwidth]{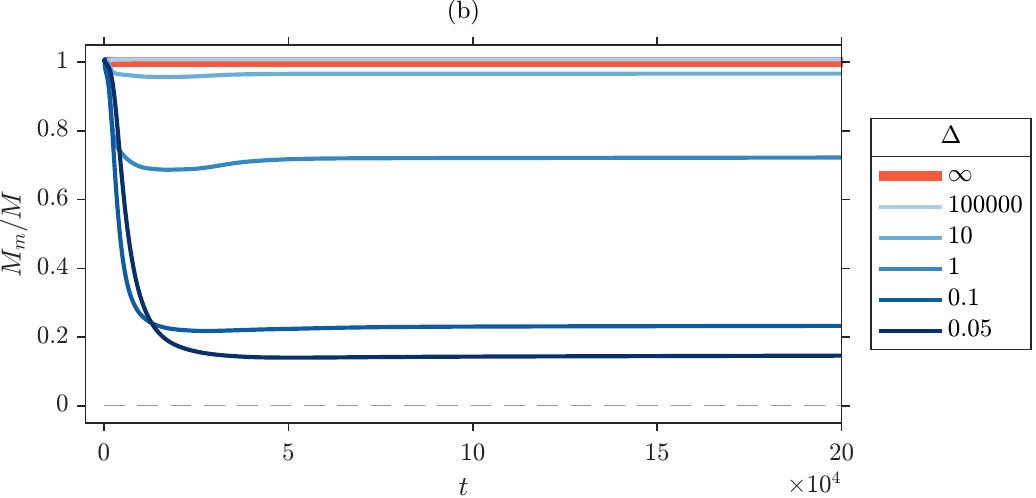}

    \caption{\label{fig:chidelta}
    Evolution of the degree of mixing ($M$) for different $\Delta$ and $\ra=10^4$ ($r=10$, see table~\ref{tab:resdisp} for further details).
    Results are shown in terms of total mixing $M$ in panel~(a), where a close up view of the early phase is also reported in the inset. 
    Here, the white symbols marks the first instant considered in the simulations.
    The initial diffusive solution~\eqref{eq:diff2} (dotted line) is also indicated.   
    The case without dispersion ($\Delta\to\infty$, red line) is shown as a reference.
    In panel~(b), the relative importance of molecular mixing to total mixing, $M_m/M$, evaluated at each instant, is shown.
    Unsurprisingly, molecular mixing becomes progressively less important as $\Delta$ increases. 
    } 
\end{figure}

The degree of mixing $M$, reported in figure~\ref{fig:chidelta}(a), confirms the previous observations. 
Initially all simulations considered follow the initial diffusive solution~\eqref{eq:diff2}.
Later, mixing in non-dispersive systems is more efficient than in cases without dispersion: the smaller the value of $\Delta$, the sooner $M$ will differ from the analytical diffusive case (inset of figure~\ref{fig:chidelta}a).
The behaviour of $M$ observed for $\Delta\ge1$ is nearly independent of $\Delta$ for long times ($t\ge1\times10^5$, main panel of figure~\ref{fig:chidelta}a), attaining $M\approx0.85$.
When $\Delta$ is further diminished, namely $\Delta\le0.1$, the evolution is impacted dramatically and the degree of mixing achieves values of 0.91 and 0.94 for $\Delta=0.05$ and $\Delta=0.01$, respectively. 
This non-monotonic behaviour suggests that a non-trivial interplay of flow evolution and onset time controls the combined dynamics of molecular and dissipative mixing.
On the one hand, the fraction of mixing due to molecular dissipation, $M_m/M$ (figure~\ref{fig:chidelta}b), increases with $\Delta$.
This is expected, since the effect of reducing $\Delta$ is to increase the importance of dispersion. 
On the other hand, the flow configuration is such that the dispersive dissipation, produced by the interplay of velocity and concentration fields, is considerably larger for $\Delta=0.1$ than for $\Delta=0.05$, see figure~\ref{fig:degdelta}(b).
This complex combination of different processes determines the evolution of the degree of mixing.
Ultimately, all cases will achieve the well-mixed condition ($M=1$).

In conclusion, present results suggest that in this configuration ($r=10,\ra=10^4$), within the range of $\Delta$ and times explored, $\Delta=0.1$ provides the most favourable mixing conditions and $M$ achieves the largest values.
The observation that it exists an optimum value of $\Delta$ that maximizes the mixing is opposite to previous findings in Rayleigh-B\'enard configuration \citep{wen2018rayleigh}, where the flux was observed to be minimum for $\Delta=0.05$.
This is not surprising, given the very different nature (transient vs. steady state, no external driving vs. constant external driving) of the two systems considered.

\section{Flow evolution with dispersion: influence of $r$}\label{sec:r}

\begin{figure}
    \centering
    \includegraphics[width=0.99\linewidth]{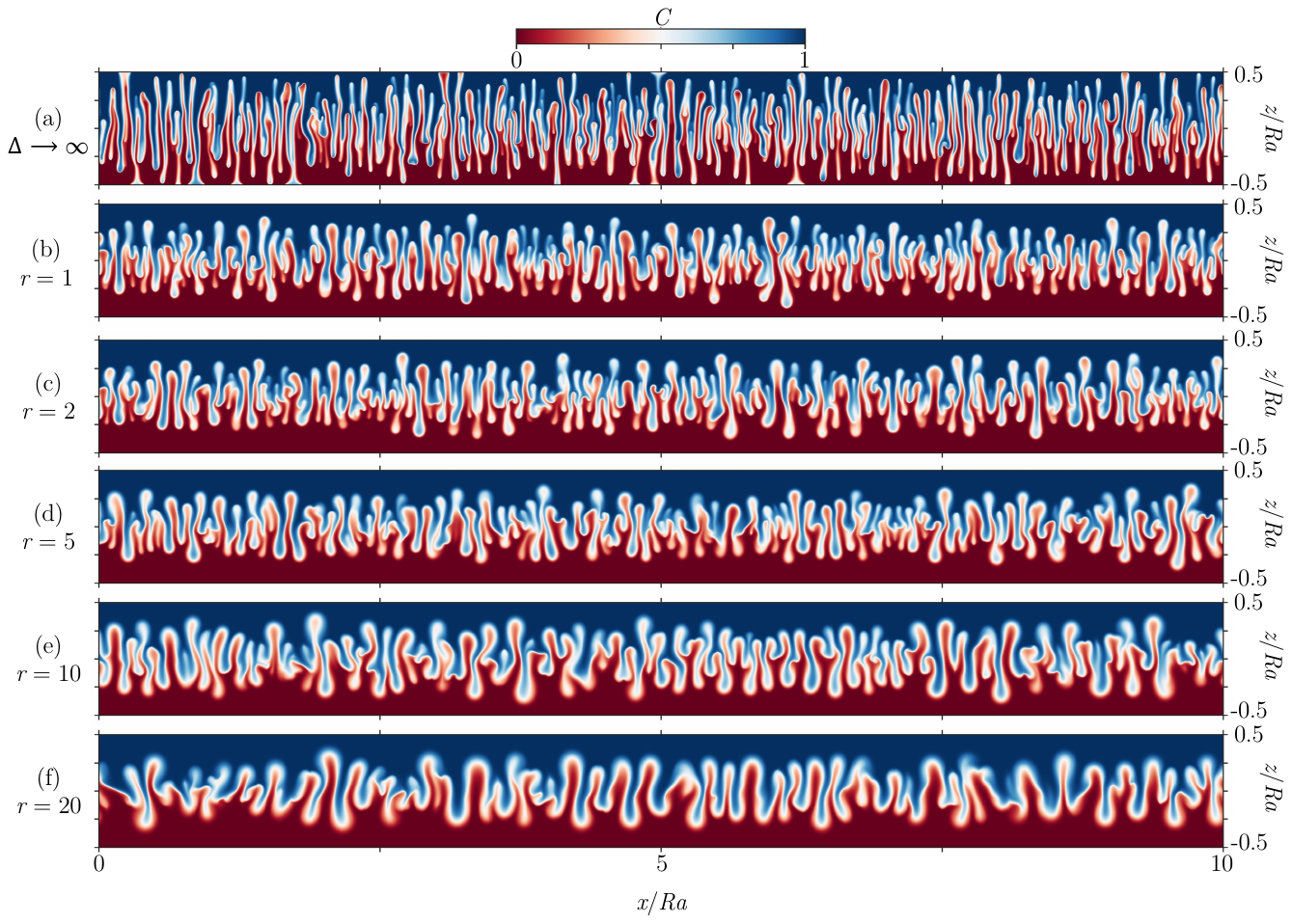}
    \caption{
    Concentration fields at $t=2\times10^4$ for different values of the dispersion parameter $r$. The field (a) corresponds to the case without dispersion ($\Delta\to\infty$).
    See Movies~\textcolor{red}{S2} and \textcolor{red}{S3} of the electronic supplementary material for the time-dependent evolution of the simulations with $r=10$ and $r=1$ (cases e and b), respectively.    
    }
    \label{fig:fieldsr}
\end{figure}

We analyse here the effect of the anisotropy ratio $r$, defined in~\eqref{eq:params}, on the flow evolution and mixing dynamics.
We fix $\Delta=0.1$, corresponding to a strongly dispersive case studied in \S\ref{sec:delta}, and we vary $r$ in the interval $1\le r \le 20$.
As previously mentioned, $r$ varies with the flow parameters, and the most common practice to consider $r\approx10$ may not be motivated, unless the problem considered is solute transport in the advection dominated regime \citep{bijeljic2007pore}.

\subsection{Flow dynamics}\label{sec:r_dyn}

The flow evolution follows the same regimes discussed in Sec.~\ref{sec:resndisp} with some differences that will be discussed in the following (see Movies~\textcolor{red}{S2} and ~\textcolor{red}{S3} of the electronic supplementary material for the time-dependent evolution of the simulations with $r=10$ and $r=1$ reported in figures~\ref{fig:fieldsdisp}e and~\ref{fig:fieldsdisp}b, respectively). 
The concentration fields taken at time $t=2\times10^4$ are reported in figures~\ref{fig:fieldsr}(b-f) for $\Delta=0.1$ and $1\le r\le 20$, and compared to the case without dispersion (figure~\ref{fig:fieldsr}a). 
We observe that, at this value of $\Delta$, the anisotropy ratio $r$ does have a role on shaping the fingers pattern: the morphology of the flow changes dramatically when $r$ is increased, corresponding to a coarsening of the fingers that become also more intricate near the centreline as $r$ rises, in agreement with previous observations by \citet{ghesmat2008viscous}.

\begin{figure}
    \centering
    \includegraphics[width=0.99\linewidth]{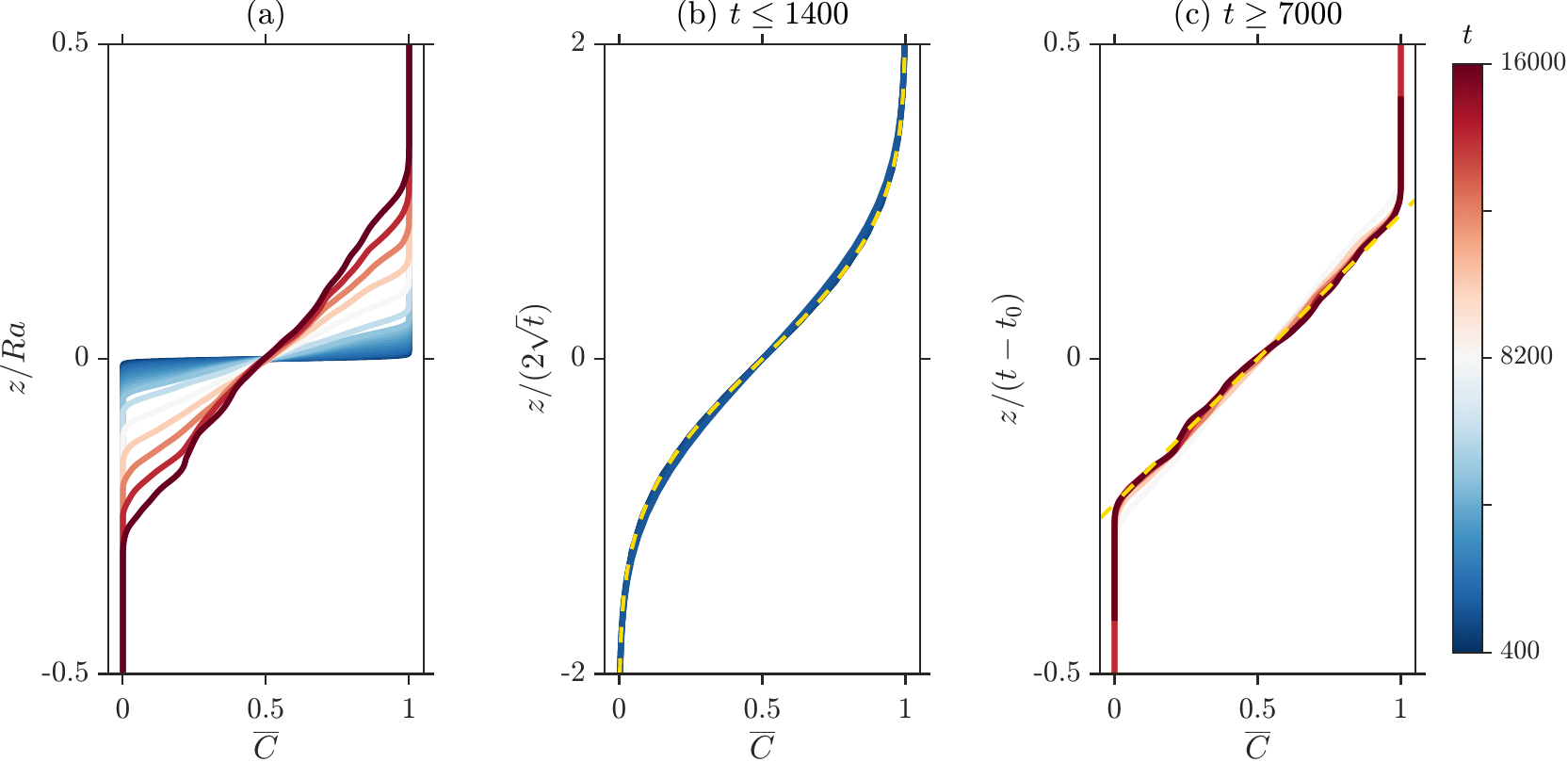} 
    \caption{Evolution of the horizontally-averaged concentration profiles, $\overline{C}$, relative to simulation $\ra=10^4$, $\Delta=0.1$ and $r=1$.
    Profiles reported correspond to instants taken preceding the fingers impact on the walls.
    Specifically, they are in the range $400\le t\le16000$ in (a), in the initial diffusive regime ($t\le1.4\times 10^3$)~in (b) and in the fingers merging and growth regime ($t\ge7\times10^3$) in (c). 
    In panel~(b), the dashed line indicates the initial diffusive solution~\eqref{eq:diffsol}.
    In panel~(c), the wall-normal coordinate is rescaled with $t-t_0$, where $t_0=4\times10^3$. 
    The dashed line represents~\eqref{eq:proffit}.
    }
    \label{fig:profiles_r}
\end{figure}

At early times ($t\le1400$), the flow dynamics is unaffected by dispersion since in this phase $\mathbf{u}\approx0$.
Therefore, the concentration field evolves according to the diffusive self-similar solution~\eqref{eq:diffsol}.
This is confirmed by the horizontally-averaged concentration profiles relative to the simulation with $r=1$ and shown in figure~\ref{fig:profiles_r}(b).
At later times, the fingers form and interact. 
The concentration fields in figure~\ref{fig:fieldsr} suggest that the concentration gradients become weaker as $r$ is increased.

We observe that, similarly to the case $r=10$ (figure~\ref{fig:profiles_delta}c), the mixing layer grows approximately linearly. 
Indeed, for $t\ge7\times10^3$, the horizontally-averaged concentration profiles of figure~\ref{fig:profiles_r}(c) are well fitted by the linear function~\eqref{eq:proffit} with $t = 4\times10^3$ and $\gamma=0.46$, where the growth rate of the mixing region, $\gamma$, matches the value obtained for $r=10$ (discussed in \S\ref{sec:delta_dyn}). 
This observation suggests that it will take a longer time for the fingers to reach the walls (corresponding to the instant at which the maximum of dissipation is achieved) compared to the case with $\Delta\to\infty$.
Following the finger's impact on the horizontal boundaries, the flow evolves towards a progressive homogenization of the concentration distribution.

\begin{figure}
    \centering
    \centering
    \includegraphics[height=0.33\columnwidth,clip=true, trim={0 0 4.5cm 0}]{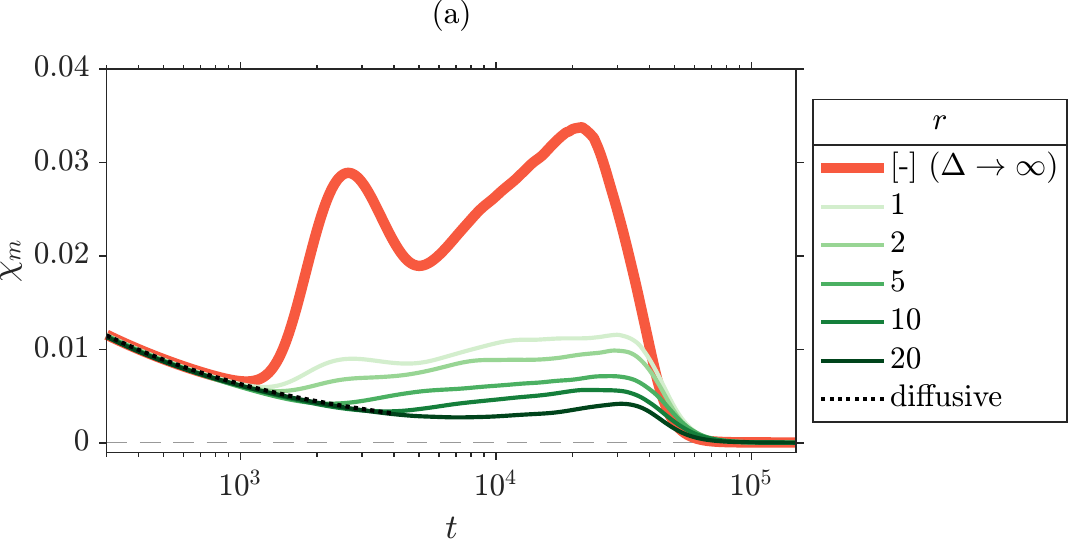}     
    \includegraphics[height=0.33\columnwidth,clip=true, trim={0 0 4.5cm 0}]{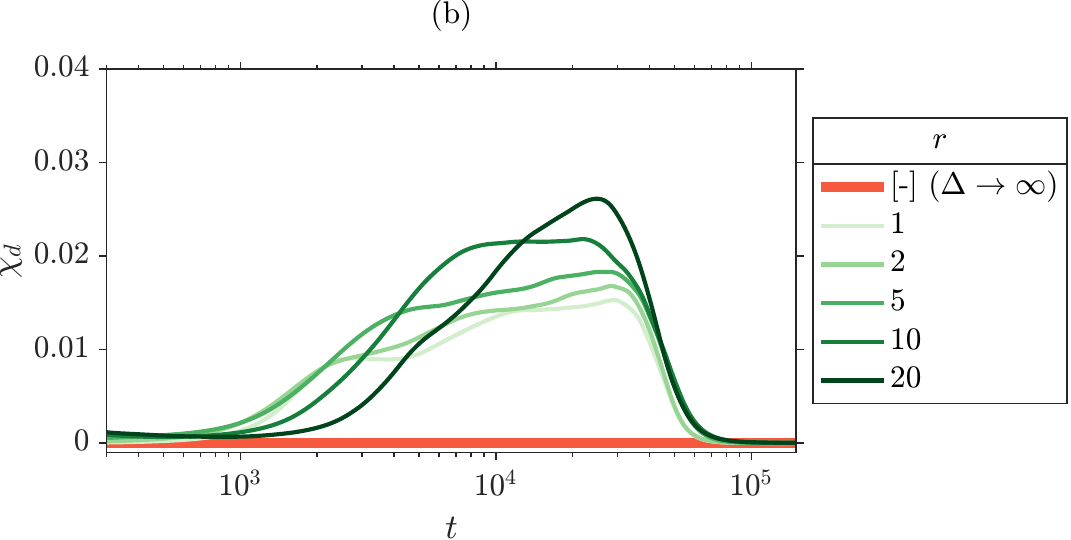}

\vspace{0.5cm}
    
    \includegraphics[height=0.33\columnwidth]{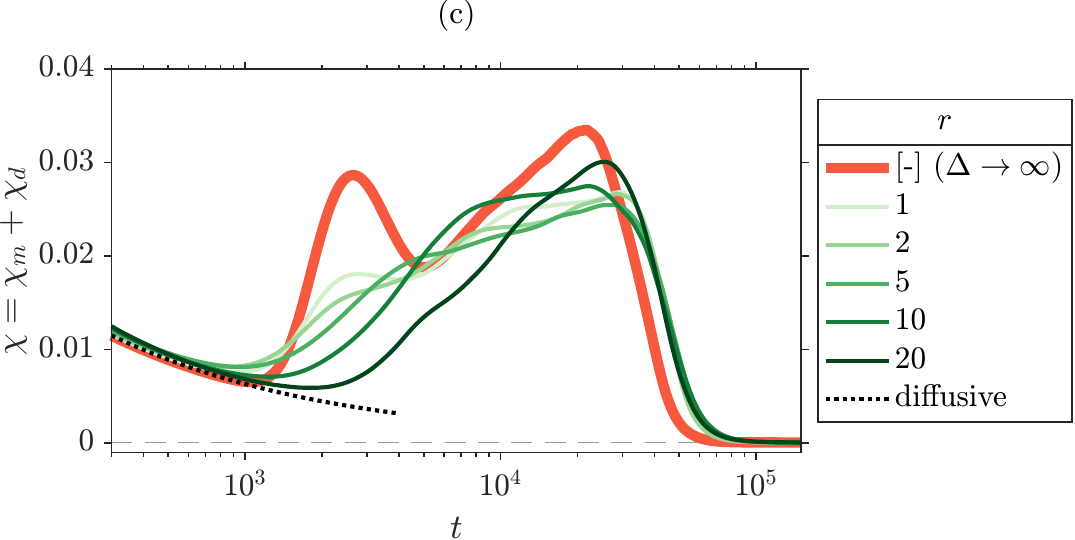}

    \caption{\label{fig:chir}
    Evolution of the mean scalar dissipation for different values of $r$.
    The red line refers to the case in absence of dispersion.
    The molecular ($\chi_m$), dispersive ($\chi_d$) and total dissipation ($\chi=\chi_m+\chi_d$) are reported in panels~(a), (b) and (c), respectively. 
    The initial diffusive solution~\eqref{eq:diff2} (dotted line) is also indicated.
    }    
\end{figure}

\subsection{Mixing}\label{sec:r_mix}

At early times, the components of the scalar dissipation reported in figure~\ref{fig:chir} follow the purely diffusive solutions~\eqref{eq:diff2}.
We observe that the smaller the value of $r$, the sooner the onset of these instabilities occurs, making the dispersive curves in figure~\ref{fig:chir}(b-c) to deviate from the  diffusive solutions. 
The onset is clearly driven by the dispersive component of the dispersion tensor: the molecular dissipation (figure~\ref{fig:chir}a) follows the diffusive solution for very long times, while the dispersive dissipation (figure~\ref{fig:chir}b) is now positive, with smaller values of $r$ leading to an earlier growth compared to larger values of $r$, in agreement with previous findings \citep{dhar2022convective}.
After formation, fingers grow vertically eventually interacting with neighbouring flow structures. 
The concentration gradients across the fingers interface reduce as $r$ is increased (figure~\ref{fig:fieldsr}), and so does the molecular component of the dissipation, $\chi_m$, as it appears from figure~\ref{fig:chir}(a).

As discussed in \S\ref{sec:r_dyn}, the fingers grow at an approximately constant rate ($\gamma=0.46$) that is lower compared to the case without dispersion ($\gamma=0.59$). 
We can infer this also from the dissipation curves in figure~\ref{fig:chir}: the maximum of each curve is achieved at nearly the same time ($t\approx3\times 10^4$).
However, the maximum value of dispersive dissipation $\chi_d$ is remarkably affected by $r$, as one would expect looking at the form of the dispersion tensor~\eqref{eq:disp01ad}.
For $\Delta=0.1$ considered here, when $t\ge 4\times 10^3$ the dispersive component of the mean scalar dissipation is always dominant compared to the molecular counterpart. 
As a result of the interplay between molecular and dispersive dissipation, the behaviour of the total dissipation $\chi$, reported in figure~\ref{fig:chir}(c), suggests that initially the mixing process is less efficient compared to the case without dispersion ($\Delta\to\infty$, red line).
In turn, for long times and after the fingers have reached the walls, $\chi$ remains higher in the cases with dispersion, with important consequences on the degree of mixing.

\begin{figure}
    \centering

\includegraphics[width=0.7\columnwidth]{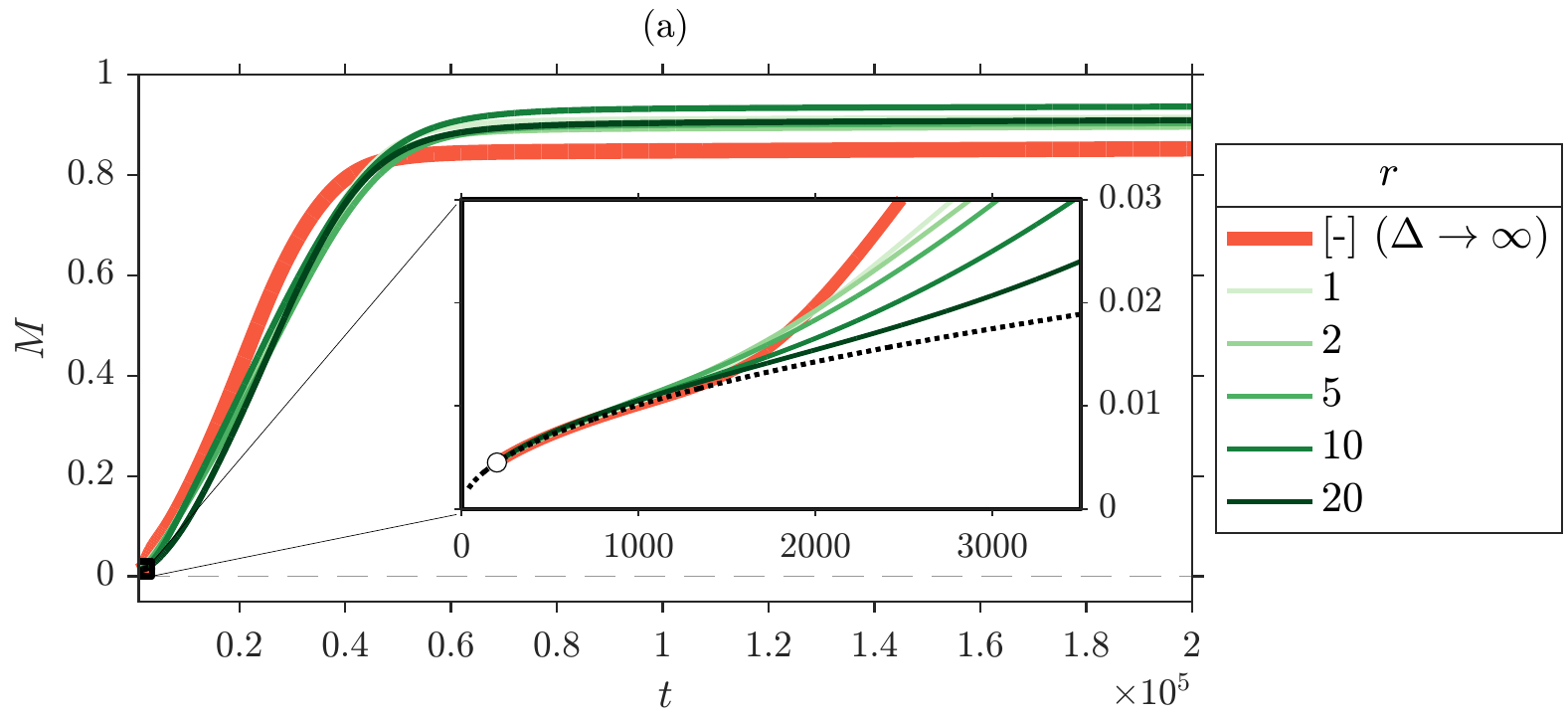} 
    
\vspace{0.5cm}
    
    \includegraphics[width=0.7\columnwidth]{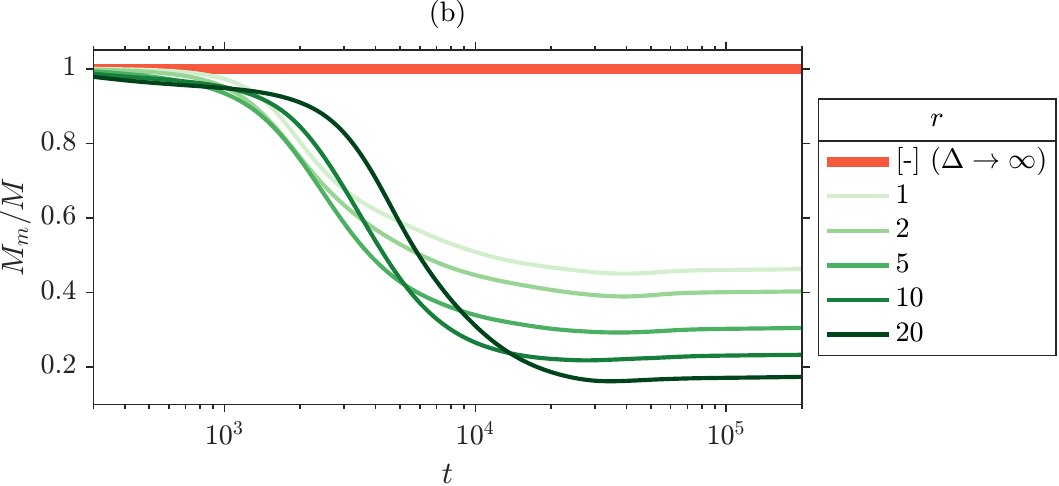}

    \caption{\label{fig:degr}
    Evolution of the degree of mixing ($M$) for different $r$ and $\ra=10^4$ ($\Delta=0.1$, see table~\ref{tab:resdisp} for further details).
    Results are shown in terms of total mixing $M$ in panel~(a), where a close up view of the early phase is also reported in the inset. 
    Here, the white symbols marks the first instant considered in the simulations.
    The initial diffusive solution~\eqref{eq:diff2} (dotted line) is also indicated.   
    The case without dispersion ($\Delta\to\infty$, red line) is shown as a reference.
    In panel~(b), the relative importance of molecular mixing to total mixing, $M_m/M$, evaluated at each instant, is shown.
    Also in this case, molecular mixing becomes progressively less important as $r$ increases. 
    }    
\end{figure}

The evolution of the degree of mixing ($M$) for different values of $r$ is shown in figure~\ref{fig:degr}.
At very early times ($t\le1500$) (see the inset of figure~\ref{fig:degr}a), the degree of mixing is larger in the simulations with dispersion, due to the earlier onset of convection previously described.
The fingers develop slower and within wider spaces compared to the case without dispersion: the gradients of concentration across the fingers interface are smaller, corresponding to values of $\chi_m$ that decrease with $r$.
Later, the dispersive mixing increases in time and also with $r$.
As a result of the interplay of these processes, the following dynamics appears: (i)~the total degree of mixing is initially large for cases with dispersion (finite $\Delta$) and low $r$; (ii)~shortly after the fingers have formed, $M$ becomes dominated by molecular diffusion and it is larger in absence of dispersion (figure~\ref{fig:degr}a, main panel, $2\times10^3\le t\le4.5\times10^4$);
(iii)~finally, after the fingers have reached the walls, $\chi_d$ decreases at a lower rate compared to $\chi_m$, and $M$ is controlled by the dispersive mixing. 
The relative importance of molecular mixing to total mixing, $M_m/M$, evaluated at each instant, is shown in figure~\ref{fig:degr}(b).
As expected, in the long term the molecular mixing becomes progressively less important as $r$ is increased, namely from 46\% to 17\% for $r=1$ and $r=20$, respectively.

\section{Conclusions and outlook}\label{sec:concl}

\subsection{Summary and conclusions}\label{sec:summconc}

We analysed the process of convective mixing in presence of dispersion in two-dimensional, homogeneous and isotropic porous media.
We considered a Rayleigh-Taylor instability in which the presence of a solute produces density differences that drive the flow.
The domain consists of two fluid layers, initially divided by a flat interface, which will mix driven by buoyancy forces and eventually reach a uniform solute distribution. 
Solute is redistributed by advection and dispersion.
The effect of dispersion is modelled using the anisotropic Fickian dispersion tensor formulation~\eqref{eq:disp01ad} proposed by \citet{bear1961tensor}.
In this model, in addition to molecular diffusion $D_m^*$, the solute is also redistributed by the spreading produced by the flow due to the complex pathways followed by the fluid parcels within the pores spaces (mechanical dispersion).
This additional solute redistribution is quantified by the longitudinal ($D_l^*$, in the direction of the flow) and the transverse ($D_t^*$, perpendicular to the direction of the flow) dispersion coefficients.
The flow is controlled by three dimensionless parameters: the Rayleigh-Darcy number $\ra$ (defining the relative strength of convection and diffusion) and the dispersion parameters $r=D_l^*/D_t^*$ and $\Delta=D_m^*/D_t^*$. 
With the aid of numerical Darcy simulations, we investigated the mixing dynamics without and with dispersion.

\begin{figure}
    \centering
\includegraphics[width=0.99\columnwidth]{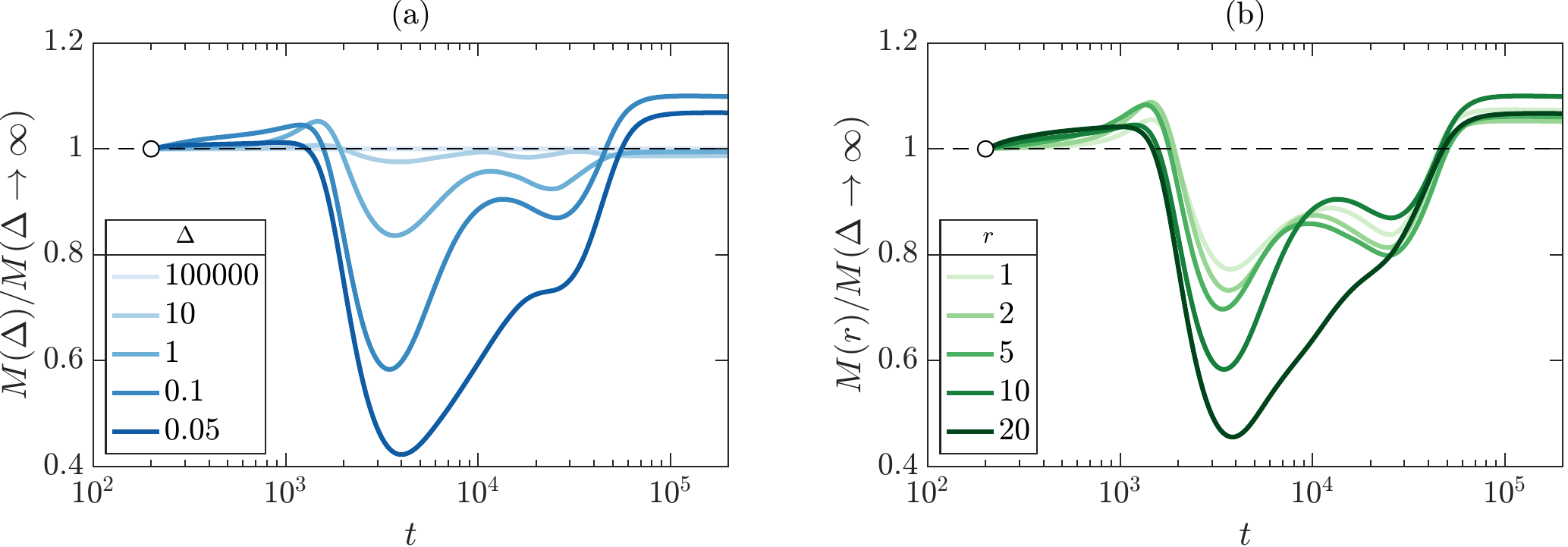} 
    \caption{\label{fig:concl}
    Influence of the dispersion parameters on the degree of mixing $M$: $\Delta$ (a) and $r$ (b). 
    Results are reported in terms of time degree of mixing $M$ relative to the case without dispersion, $M(\Delta\to\infty)$.
    The first instant considered in the simulations (bullet) is also indicated.
    }    
\end{figure}

We found that in absence of dispersion ($\Delta\to\infty$) the system's dynamics is self-similar and independent of $\ra$ until the fingers approach the domain's horizontal boundaries, and the flow evolution follows several regimes, which we describe.
Then we analysed the effect of dispersion in time ($t$) for a fixed value of Rayleigh number ($\ra=10^4$).
We quantify the mixing state of the system using the degree of mixing $M$ \citep{jha2011quantifying}, which varies between $M(t=0)=0$ (segregated layers) and $M(t\to\infty)=1$ (uniform solute concentration field).
A detailed analysis of the scalar dissipation reveals a complex interplay between flow structures and mixing due to the dispersive and molecular contributions.
Both $r$ and $\Delta$ affect the fingers formation and development: in particular, the lower the value of $\Delta$ (or the larger the value of $r$), the wider, the more convoluted and diffused the fingers (see figures~\ref{fig:fieldsdisp} and \ref{fig:fieldsr}).
For an anisotropic dispersion tensor with $r=10$ \citep[representative of solute transport in the advection dominated regime,][]{bijeljic2007pore}, the role played by the relative importance of molecular and transverse dispersion, $\Delta$, is crucial.
This is presented in figure~\ref{fig:concl}(a) in terms of degree of mixing, $M(\Delta)$, normalized by the case without dispersion $M(\Delta\to\infty)$. Three main phases appear: (i)~initially ($t\le2\times10^3$) the mixing is more efficient in cases with dispersion; 
(ii)~after the fingers have grown sufficiently and merged, the mixing performance in the dispersive cases is lower than in absence of dispersion; 
(iii)~for longer times, in contrast, high dispersion flows ($\Delta\le0.1$) show higher degree of mixing compared to systems without dispersion.
Ultimately, all cases will lead to the same uniform configuration ($M=1$).
Remarkably, we found that within the time frame considered, it exists an optimum value $\Delta=0.1$ that maximizes the mixing. 
We finally looked at the impact of $r$ on the mixing when $\Delta =0.1$ (figure~\ref{fig:concl}b).
We found that, for such value of $\Delta$, the anisotropy ratio $r$ produces only second order effects, with some noticeable changes only in the intermediate phase, where it appears that the mixing is more efficient for small values of anisotropy.

\subsection{Implications for geophysical flows: saline seepage in groundwater systems}\label{sec:summimpl}
The theoretical framework proposed allowed us to analyse numerical Darcy simulations with dispersion by splitting the mean scalar dissipation into molecular and dispersive components.
In combination with pore-scale simulations and bead packs experiments, this approach can be used to validate and improve current dispersion models to obtain more reliable estimates of solute transport and spreading in the subsurface.
As an example of a possible application, we discuss here the case of saline seepage from salt water basins presented by \citet{narayan1995simulation}, and summarized in the following. 

\begin{figure}
    \centering
    \includegraphics[width=\linewidth]{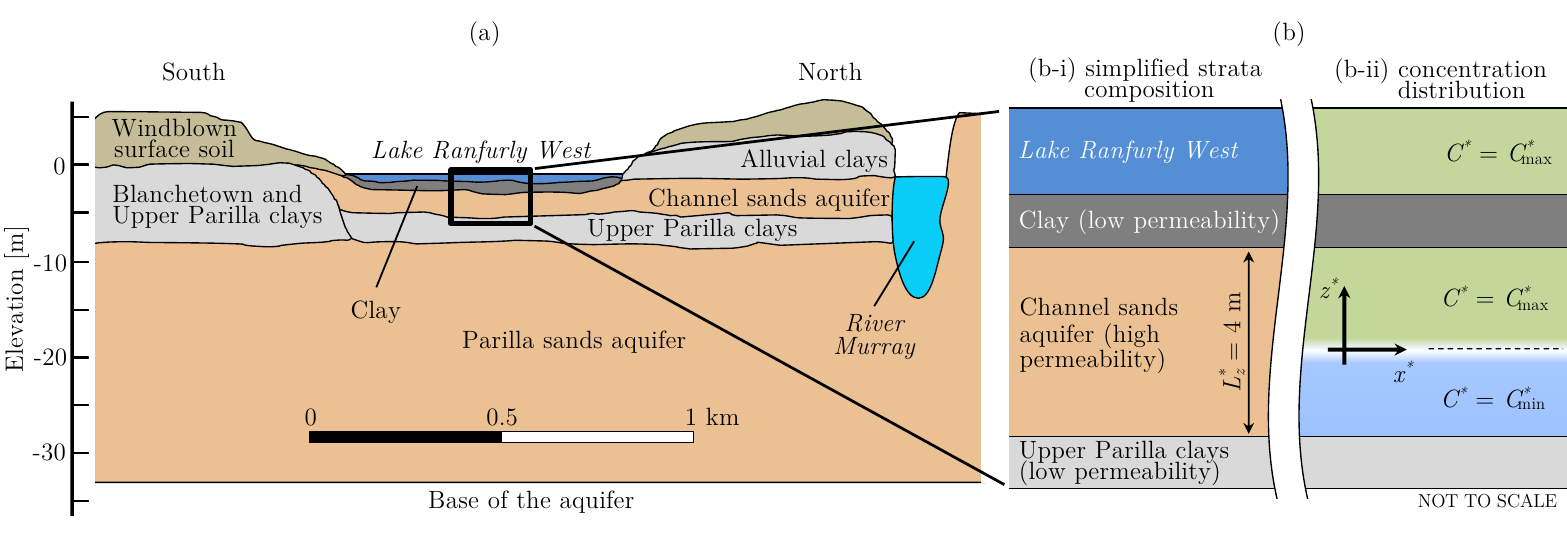}
    \caption{
    (a)~Conceptualized hydrogeology of the River Murray basin area, adapted from \citet{narayan1995simulation}.
    (b)~Modelling of the saline seepage through the bottom of the Lake Ranfurly West.
    (b-i)~The high-permeability sands aquifer is confined by two low-permeability layers. 
    (b-ii)~The Lake Ranfurly West supplies high salt concentration water ($C^*=C^*_\text{max}$) from the top, while low salinity ($C^*=C^*_\text{min}$) groundwater is present in the aquifer.
    }
    \label{fig:appli}
\end{figure}

One of the most important drainage systems in Australia is represented by the Murray-Darling river, a key source of water in the region.
Near-surface groundwater in this basin is characterized by high salinity.
Agricultural activities have led to a rising of the water tables, with the consequence of an increased discharge of high salt concentration groundwater into the river basin.
This process may eventually increase the river's water salinity to unacceptable levels in periods of low flow rate in the river.
To prevent this issue, high-salinity groundwater is intercepted and stored in basins at the surface level, where it may evaporate, further increasing the salt concentration.
In some cases, these surface basins are designed to allow a slow and controlled leak to the underlying Murray River aquifer.
This is the case for the Lake Ranfurly West \citep{ghassemi1988effect}, which releases high-salinity water through the Channel sands aquifer into the River Murray.
The hydrogeology of the system is sketched in figure~\ref{fig:appli}(a).
Designing and controlling such basins is key to manage the water resources efficiently and to keep the salinity of the rivers at an acceptable level. 
Here we will apply our findings to determine the role of dispersion in the salt spreading process from the Lake Ranfurly West and the River Murray basin.

The flow can be modelled as a high permeability region (Channel sands aquifer) confined by two low-permeability clay layers (see figure~\ref{fig:appli}b-i): the clay layer at the bottom of the lake confines the system from above, and the Upper Parilla clays confine the system from below \citep{narayan1995simulation}.
The fluid saturating the high-permeability region can be modelled as an heavy and high-salt concentration fluid ($C^*=C^*_\text{max}=0.120$~kg/kg, seeping from the Lake Ranfurly West) sitting on top of a lighter, low salinity fluid ($C^*=C^*_\text{min}=0.045$~kg/kg, groundwater present in the Channel sands aquifer).
A simplified flow configuration associated with this system is illustrated is figure~\ref{fig:appli}(b). 
We neglect here any flow circulations induced by pumping, and we assume that the flow is only driven by density differences. 
We take the system parameters from \citet{narayan1995simulation} and determine the relevant dimensionless flow quantities as follows. 
The depth of the porous layer is $L_z^*=4$~m, with porosity $\phi=0.3$ and permeability $k=2.95\times10^{-11}$~m$^2$ (we consider the medium isotropic).
The salt concentration difference $C^*_\text{max}-C^*_\text{min}$ produces a maximum density difference $\Delta\rho^*=52.5$~kg/m$^3$.
Considering a dynamic viscosity $\mu=10^{-3}$~Pa~s, the characteristic velocity is $\mathcal{U^*}=1.52\times10^{-5}$~m/s~$=1.31$~m/d, and we obtain a Rayleigh number $\ra=1.35\times10^5$.
The longitudinal dispersivity is $\alpha^*_l = 80$~m and the anisotropy ratio $r=10$.
With this set of parameters, we have $\Delta = rD_m/(\mathcal{U}^*\alpha^*_l)\approx10^{-5}$.
From the results presented in figure~\ref{fig:concl}(a) we conclude that dispersion is the dominant mixing mechanisms, and has to be accounted in the design and simulation of these flow. 
We also remark that, despite the high uncertainty on $\alpha^*_l$, also much smaller values would lead to the same conclusion, e.g., for $\alpha^*_l = 0.8$~m we have that $\Delta \approx10^{-3}$, still in the range of parameters in whcih dispersion represents the dominant mixing mechanism. 
This would change only if substantially lower values of $\ra$ are considered.

\subsection{Limitations and future developments}\label{sec:limout}
A possible limitation of the present work consists of the model adopted. 
Despite being widely employed \citep{depaoli2023review}, the dispersion model considered is derived for idealized advective flows, i.e., where the source of driving is a pressure gradient that generates a constant and uniform velocity field.
However, the configuration studied here is transient, and therefore the flow velocity scales also evolve across the different regimes experienced by the flow.
An interesting possible development consists of taking the variation of the velocity scales into account in the dispersion model, to better describe the dispersion coefficients $D_l^*$ and $D_t^*$ \citep{koch1985dispersion,delgado2007longitudinal}. 
In addition, recently a theory that quantifies the interplay between intrapore and interpore flow variabilities and their impact on hydrodynamic dispersion has been proposed \citep{liu2024scaling}.
Extending the model of \citet{liu2024scaling} to explore buoyancy-driven mixing would improve the reliability of the results obtained via numerical simulations in the present context \citep{Woods2025}.

The dimensionality of the system will likely have an effect on the mixing process, as previously observed by \citet{Boffetta2020} and \citet{boffetta2022dimensional}, who studied the present flow configuration without dispersion.
In particular, they have shown that the growth of the mixing layer is strongly affected by the dimensionality of the domain, with the fingers being more coherent in 2D than in 3D. This leads to a faster growth of the mixing region and to a larger variance of the concentration distribution in the 2D case.
The effects of dimensionality have been recently explored in the one-sided configuration \citep{depaoli2025grl}, and the same flow dynamics has been observed in 2D and in 3D, with a larger fingers size and a lower mixing rate in two dimensions. 
In contrast, still in one-sided configuration but in presence of dispersion, \citet{dhar2022convective} observed that the fingers size is independent of the dimensionality at $\ra=10^3$, while it is larger in 3D than in 2D at $\ra=3\times10^3$, and in all cases the fingers grow at the same rate.
They also observed that the effect of dispersion on the onset time and on the mixing are similar in 2D and in 3D. 
Considering these findings, we expect that in a three-dimensional, porous and dispersive Rayleigh-Taylor flow, the mixing dynamics will exhibit: (i)~a slower growth rate of the mixing region, (ii)~wider fingers, and (iii)~similar effects of dispersion on the mixing dynamics, compared to the two-dimensional case presented here.
To confirm these assumptions, a full campaign of 3D numerical simulations is required.

In this work, we considered fully miscible fluids.
In geological processes, however, the presence of partially miscible fluids (e.g., CO$_2$ and brine) may significantly affect the flow dynamics and solute mixing processes.
With the aid of a phase field method (Darcy-Cahn-Hilliard), \citet{li2022diffuse,li2023dissolution} have recently shown that, in semi-infinite domains, partial miscibility of the fluids leads to larger dissolution fluxes compared to the case of fully miscible fluids, corresponding to a more efficient mixing process.
We believe that in the present configuration the same effect could possibly occur at early times, when the driving, provided by the volume of fresh fluid available outside the mixing region, is nearly constant.

Finally, our findings have been obtained in idealized conditions, since we considered homogeneous and isotropic porous domains. 
However, geological formations are heterogeneous at multiple scales, requiring Darcy simulations to include non-uniformities in the permeability and porosity fields \citep{woods2015flow}.
An interaction between the scales of the heterogeneity and of the instability that controls the flow evolution has been observed \citep{benhammadi2025experimental}.
Heterogeneities can also influence the form of the dispersion tensor, and possibly lead to non-Fickian effects \citep{cala1986velocity}.
A systematic study of Rayleigh-Taylor flows in heterogenous porous domains at the Darcy scale, combined to pore-scale simulations and experiments, is advisable to shed new light on the role of heterogeneities, and eventually to improve and validate existing dispersion models.

\appendix
\section*{Appendix}

\section{Additional details on the dispersion tensor}\label{sec:appB}
In a three-dimensional reference frame, the dispersion tensor~\eqref{eq:disp01ad} with all the terms written explicitly reads \citep{zheng1999mt3dms}:
\begin{equation}
\mathsfbi{D}=
\begin{bmatrix}
1+\frac{1}{\Delta}\left[\lvert\mathbf{u}\rvert+(r-1)\frac{u^{2}}{\lvert\mathbf{u}\rvert}\right] & 
(r-1)\frac{uv}{\Delta\lvert\mathbf{u}\rvert} & 
(r-1)\frac{uw}{\Delta\lvert\mathbf{u}\rvert} \\
(r-1)\frac{uv}{\Delta\lvert\mathbf{u}\rvert} & 1+\frac{1}{\Delta}\left[\lvert\mathbf{u}\rvert+(r-1)\frac{v^{2}}{\lvert\mathbf{u}\rvert}\right]  & (r-1)\frac{vw}{\Delta\lvert\mathbf{u}\rvert} \\
(r-1)\frac{uw}{\Delta\lvert\mathbf{u}\rvert} & (r-1)\frac{vw}{\Delta\lvert\mathbf{u}\rvert} & 1+\frac{1}{\Delta}\left[\lvert\mathbf{u}\rvert+(r-1)\frac{w^{2}}{\lvert\mathbf{u}\rvert}\right]  \\
\end{bmatrix}.
\label{eq:disp01b}
\end{equation}
Note that expanding the divergence term of the right hand side of~\eqref{eq:equ1bis1} we obtain: 
\begin{align}
\nabla\cdot(\mathsfbi{D}\nabla C) &= \frac{\partial }{\partial x_{i}}\left( D_{ij}\frac{\partial C}{\partial x_{j}}\right)=
\frac{\partial D_{ij}}{\partial x_{i}}\frac{\partial C}{\partial x_{j}}+ D_{ij}\frac{\partial }{\partial x_{i}}\left(\frac{\partial C}{\partial x_{j}}\right)\\
\begin{split}
&= D_{xx}\frac{\partial^{2} C}{\partial x^{2}}+D_{yy}\frac{\partial^{2} C}{\partial y^{2}}+D_{zz}\frac{\partial^{2} C}{\partial z^{2}}+
\frac{\partial C}{\partial x} \left(\frac{\partial D_{xx}}{\partial x}+\frac{\partial D_{xy}}{\partial y}+\frac{\partial D_{xz}}{\partial z}\right)+\\
&\quad+ \frac{\partial C}{\partial y} \left(\frac{\partial D_{xy}}{\partial x}+\frac{\partial D_{yy}}{\partial y}+\frac{\partial D_{yz}}{\partial z}\right)+
 \frac{\partial C}{\partial z} \left(\frac{\partial D_{xz}}{\partial x}+\frac{\partial D_{yz}}{\partial y}+\frac{\partial D_{zz}}{\partial z}\right)+\\
&\quad+2\left(
D_{xy}\frac{\partial^{2} C}{\partial x\partial y}
+D_{yz}\frac{\partial^{2} C}{\partial y\partial z}
+D_{xz}\frac{\partial^{2} C}{\partial x\partial z}\right).
\end{split}
\label{eq:disp06}
\end{align}
The right hand side of~\eqref{eq:disp06} reduces to the Laplacian form $D_m^*\nabla^{2}C$ when $D_{ii}=D_m^*$ (with $D_m^*$ uniform in space) and $D_{ij, i\ne j}=0$.
If the dispersion tensor~\eqref{eq:disp01ad} is isotropic ($r=1$, see~\S\ref{sec:r}), then~\eqref{eq:disp06} takes the form
\begin{align}
\begin{split}
\nabla\cdot(\mathsfbi{D}\nabla C)  
&= \frac{\partial }{\partial x_i}\left(D_{ii}\frac{\partial C}{\partial x_i}\right)\\
&=\frac{\partial C}{\partial x} \frac{\partial D_{xx}}{\partial x}+ \frac{\partial C}{\partial y} \frac{\partial D_{yy}}{\partial y}+ \frac{\partial C}{\partial z} \frac{\partial D_{zz}}{\partial z}+D_{xx}\frac{\partial^{2} C}{\partial x^{2}}+D_{yy}\frac{\partial^{2} C}{\partial y^{2}}+D_{zz}\frac{\partial^{2} C}{\partial z^{2}},
\end{split}
\label{eq:disp07}
\end{align}
where the convention for summation with respect to repeated indices is used.
In case the coefficients $D_{ii}$ are constant, the Laplacian form is again recovered.

\section{Numerical details}\label{sec:appA}
\subsection{Numerical treatment of the dispersive term}\label{sec:appA1}
A possible discretization of equation~\eqref{eq:equ1bis1} consists of splitting the terms as follows:
\begin{equation}
\label{eq:dispnum1}
\underbrace{\left(\frac{\partial}{\partial t} - \nabla^2\right)C}_{\text{Implicit terms}} = \underbrace{\nabla \cdot \left([\left(\mathsfbi{D}-\mathsfbi{I}\right) \nabla C\right] - \mathbf{u}\cdot\nabla C}_{\text{Explicit terms}}.
\end{equation}
In this method, the term on the left hand side of the equation is treated implicitly and the term on the right hand side is treated explicitly~\citep[the explicit treatment of the dispersive term was also employed by][]{wen2018rayleigh}. However, this approach is very restrictive on the size of the time step. To ensure numerical stability, the time step must scale with the square of the minimum grid spacing. For finely resolved boundary layers, this makes the computation unrealistically expensive. 
In order to remedy this issue, the following strategy is employed: 
(i)~first, a predictor-corrector type approach is used with \eqref{eq:dispnum1} serving as the predictor step, in order to obtain the intermediate solution for the concentration field $C^{n+1/2}$, with the superscripts $n$ indicating current time step; 
(ii)~then equations~\eqref{eq:equ1bis2} and \eqref{eq:equ1bis3} are solved to obtain an intermediate solution for the velocity field $\mathbf{u}^{n+1/2}$; 
(iii)~next, the dispersion coefficients for $\mathsfbi{D}^{n+1/2}-\mathsfbi{I}$ are computed using the intermediate velocity field;  
(iv) finally, the corrector step is performed by treating only the advective terms $\mathbf{u}^n\cdot\nabla C^n$ explicitly, while the rest of the terms can be treated implicitly as 
\begin{equation}
\label{eq:dispnum2}
\underbrace{\left[\frac{\partial}{\partial t} - \nabla \cdot \left(\mathsfbi{D}\nabla\right)\right]C}_{\text{Implicit terms}} = \underbrace{ - \mathbf{u}\cdot\nabla C}_{\text{Explicit terms}}.
\end{equation}
Traditionally, the discretized form of equation~\eqref{eq:dispnum1} could be solved in AFiD \citep{depaoli2025afid} by factorizing the implicit terms effectively splitting the corresponding discrete operator into three discrete linear operators along each of the Cartesian dimensions. Such an algorithm cannot be used to solve the discretized form of equation~\eqref{eq:dispnum2}, and therefore we use iterative methods.
Since it is difficult to realize highly scalable parallel implementations of methods with fast convergence (e.g., Gauss-Seidel), the Jacobi iteration method \citep{ferziger2019computational} is used.
For sufficiently large $\ra$, the discretized operator matrix resulting from the implicit terms is strictly diagonally dominant, thereby guaranteeing convergence. Implicitly treating the dispersion terms in this way ensures the numerical stability of the solution for larger time steps. In this case, the size of the time step is only restricted by the CFL number determined by the advective terms.

\begin{figure}
    \centering
    \includegraphics[width=0.6\linewidth]{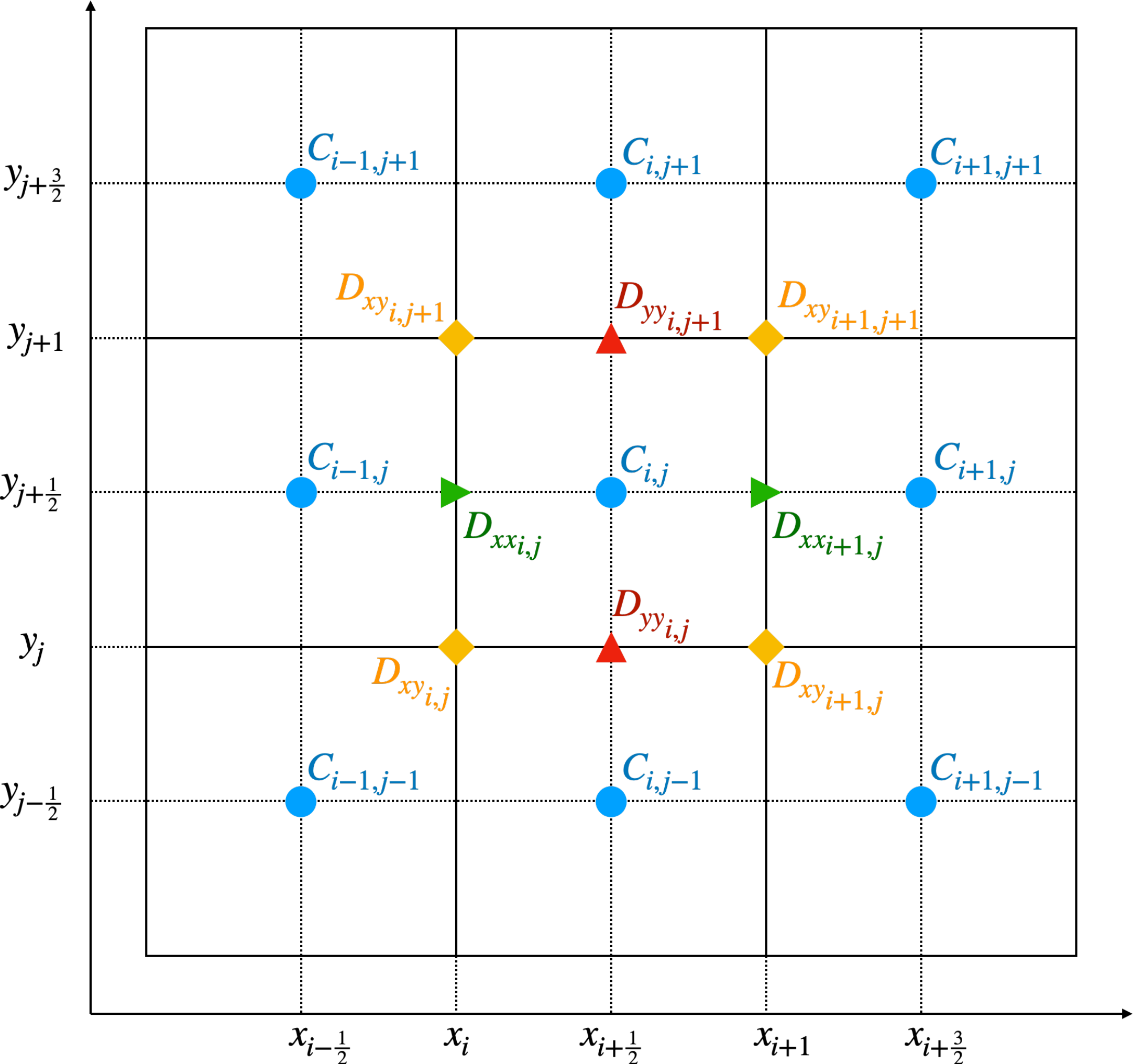}
    \caption{Grid layout for demonstrating the computation of dispersion terms in two dimensions. Grid points containing information of variable $C$ are indicated in blue, $D_{xx}$ in green, $D_{yy}$ in red, and $D_{xy}$ in yellow. Note that the subscripts $i,j$ here no longer refer to the indices of the general dispersion tensor $\mathsfbi{D}$, instead indicating the grid coordinates along $x$ and $y$ axes, respectively.
    Additional details on the variables arrangement on the grid are provided by \citet{depaoli2025afid}.}
    \label{fig:dispgrid}
\end{figure}

In the following, the discretization of the dispersive terms is explained by using a two-dimensional example in the $x,y$ space, but it can easily be extended to three dimensions. Consider a staggered grid as shown in figure~\ref{fig:dispgrid} with grid points containing the information of the concentration field $C$ at the cell-centres. 
We use linear interpolations of the velocity components to compute $D_{xx}$ at face-centres where $u$ is stored, $D_{xy}$ at cell-vertices, and $D_{yy}$ at face-centres where $v$ is stored \citep[additional details on the variables arrangement on the grid are provided by][]{depaoli2025afid}.
For the predictor step, $D_{xx} - 1$ is used instead of $D_{xx}$, $D_{yy} - 1$ is used instead of $D_{yy}$, and $D_{zz} - 1$ is used instead of $D_{zz}$. For the corrector step, we compute and store the coefficients of the 26 neighbouring concentration grid points in a stencil that spans 3 grid cells along each dimension. These coefficients are used as matrix elements in the Jacobi iteration method.

\subsection{Initial condition and grid resolution}\label{sec:appA2}
The initial condition is determined from the analytical solution~\eqref{eq:diffsol} of the advection-dispersion equation~\eqref{eq:equ1bis1}, discussed in \S\ref{sec:resndisp}.
In correspondence of the time ($t_0$) employed to initialize the flow, the initial concentration distribution is
\begin{equation}
C(z,t=t_0)=\frac{1}{2}\left[1+\text{erf}\left(\frac{z}{2\sqrt{t_0}}\right)\right].
\label{eq:b3}
\end{equation}
The inverse function associating $\tilde{z}$ to a given value of concentration $\tilde{C}$ at time $t_0$ is:
\begin{equation}
\tilde{z}(\tilde{C},t=t_0)=2\sqrt{t_0}\text{ erf}^{-1}(2\tilde{C}-1)
\label{eq:appb1}
\end{equation}
and allows to determine the minimum requirements in terms of combination of grid resolution/initial time considered.
Assuming we aim at solving from 1\% to 99\% of the initial scalar difference with at least $n_z$ points and the grid used is uniform, using~\eqref{eq:appb1} we have that the grid spacing in wall-normal direction is:
\begin{equation}
\Delta z = \frac{\ra}{N_z-1} \le \frac{\tilde{z}_{99}-\tilde{z}_1}{n_z}
= \frac{2\sqrt{ t_0}}{n_z}\left[\text{erf}^{-1}(0.98)-\text{erf}^{-1}(-0.98)\right] 
\approx 6.580\frac{\sqrt{t_0}}{n_z}.
\end{equation}
Therefore, to resolve the 1\% to 99\% scalar variation with at least $n_z$ grid points we need to initialize the flow with $t_0\ge 0.0231\times [n_z\ra/(N_z-1)]^2$.
Using $N_z=1024$ and $\ra=10^4$, it gives $t_0\approx20$ ($n_z=3$) and $t_0\approx80$ ($n_z=6$).
All simulations are initialized with the same initial perturbation at $t_0=50$.
The simulations are initially run without dispersion up to $t=200$, and the dispersive terms are later activated.
Note that $t=200$ falls well within the diffusive part of the simulations (see, e.g., figure~\ref{fig:degdelta}a), and therefore this formulation does not affect the subsequent formation and development of the flow structures.
In addition, this strategy ensures that the same initial concentration and velocity fields are employed in all the simulations, making the different behaviour observed for $r$ and $\Delta$ independent of the initial flow configuration, and also providing a smooth initial condition to the simulations with dispersion.
We wish to remark that minor discrepancies in the diffusive regime compared to the theoretical solution (see $\chi_d$ at early times and for high dispersion, figures~\ref{fig:degdelta}b and~\ref{fig:chir}b) are due to the resolution, and doubling the resolution would resolve this minor issue.

\section{Analytical solution for diffusion in confined domains}\label{sec:appCall}
\subsection{Initial diffusion at low Rayleigh number}\label{sec:appC}
We derive here the analytical solution for the mean scalar dissipation assuming pure diffusion in a confined domain with with no-flux boundary conditions.
We use~\eqref{eq:equ1bis1}, we assuming that the fluid is still ($\mathbf{u}=0$) and consider the system uniform in horizontal direction ($\partial_x C = 0$).
In addition, we assume that this regime starts at $t=t_i$.
Introducing the rescaled dimensionless coordinate $\bar{z} = z + \ra/2$, the exact solution is \citep{strauss2007partial}:
\begin{equation}
C(\bar{z},t) = \sum_{n=0}^\infty A_n \exp{\left[-\left(\frac{n\pi}{\ra}\right)^2t\right]}\cos{\left(\frac{n\pi \bar{z}}{\ra}\right)}\text{  .}
\label{eq:ddr1cd}
\end{equation}
The coefficients of the cosine series~\eqref{eq:ddr1cd} are defined as:
\begin{equation}
    A_n = 
    \begin{cases}
        \displaystyle\frac{1}{\ra}\int_0^{\ra} C(\bar{z},t_i) \textrm{ d}\bar{z} \qquad,\quad\text{for } n=0 \\
        \displaystyle\frac{2}{\ra}\int_0^{\ra} C(\bar{z},t_i) \cos{(n\pi \bar{z})} \textrm{ d}\bar{z} \qquad,\quad\text{for } n\ge1
    \end{cases}
\label{eq:ddr2bcd}
\end{equation}
with $C(\bar{z},t_i)$ defined as the initial condition~\eqref{eq:b3}, that in terms of $\bar{z}$ reads:
\begin{equation}
    C(\bar{z},t_i) = \frac{1}{2}\left[1+\text{erf}\left(\frac{\bar{z}-\ra/2}{2\sqrt{t_i}}\right)\right].
\label{eq:icfouriercd}
\end{equation}
Using the definition of molecular mean scalar dissipation~\eqref{eq:defdiss} and the analytical diffusive solution~\eqref{eq:ddr1cd}, the mean scalar dissipation is: 
\begin{align}
\chi_m(t+t_i)&=\int_{0}^{\ra}\left(\frac{\partial C}{\partial \bar{z}}\right)^2 \text{ d}\bar{z}\\
&=\int_0^{\ra}\left(\sum_{n=0}^\infty 
\frac{n\pi}{\ra} A_n 
\exp{\left[-\left(\frac{n\pi}{\ra}\right)^2t\right]}\sin{\left(\frac{n\pi \bar{z}}{\ra}\right)}\right)^2\text{ d}\bar{z}.
\label{eq:ref39appcd}
\end{align}
The solution \eqref{eq:ref39appcd} for $\ra=10^2$ is shown in figure~\ref{fig:chira}.
It has been computed using $n=100$, with a spatial discretization of 128 points for $\bar{z}$ and $t_i=50$, i.e. the same grid size and initial time used in the corresponding simulation.
Note the time shift $t_i$ in the solution~\eqref{eq:ref39appcd}, due to the fact that~\eqref{eq:ddr1cd} is computed assuming $t=t_i$ as the initial time.
The accuracy of the analytical solution in predicting the decay of dissipation obtained in the simulation is excellent. 
In addition, we observe in~\eqref{eq:ref39appcd} that $\partial C/\partial \bar{z}\sim\exp(-n^2)$, indicating a fast decay with $n$.
We verified that $n=1$ is sufficient to capture well the decay, and additional terms improve the behaviour only for the very early times $(t\le60)$.

\subsection{Final diffusion at high Rayleigh numbers}\label{sec:appC2}
As discussed in \S\ref{sec:resndisp}, the evolution during the final diffusive regime starts at time $t_f$ with a linear concentration profile: 
\begin{equation}
    C(\bar{z},t_f) = \frac{1}{2} + \beta\left(\frac{1}{2} - \frac{\bar{z}}{\ra}\right).
\label{eq:icfouriercd2}
\end{equation}
The initial condition is set with $\beta=0.25$ and $t_f=2.5\times10^4$ for $\ra=10^3$ and $\beta=0.70$ and $t_f=10^5$ for $\ra=10^4$, and it is used to determine the cosine series coefficients as in~\eqref{eq:ddr2bcd}.
Finally, the evolution of the mean scalar dissipation is determined using~\eqref{eq:ref39appcd} as:
\begin{equation}
\chi_m(t+t_f) = \int_0^{\ra}\left(\sum_{n=0}^\infty 
\frac{n\pi}{\ra} A_n 
\exp{\left[-\left(\frac{n\pi}{\ra}\right)^2t\right]}\sin{\left(\frac{n\pi \bar{z}}{\ra}\right)}\right)^2\text{ d}\bar{z},
\label{eq:ref39appcd2}
\end{equation}
and the results are reported in figure~\ref{fig:darcydr}, where a spatial discretization of 512 points for $\bar{z}$ is used.

\backsection[Acknowledgements]{Baole Wen is gratefully acknowledged for providing the data of \citet{wen2018rayleigh}, used to validate the code with dispersion.
Jeff Wood, Yantao Yang and Chenglong Hu are also acknowledged for useful discussions.
Three anonymous Referees are also acknowledged for their insightful comments and suggestions.
} 

\backsection[Funding]{
This project has received funding from the European Union's Horizon Europe research and innovation programme under the Marie Sklodowska-Curie grant agreement MEDIA No.~101062123.
We acknowledge the EuroHPC Joint Undertaking for awarding the projects EHPC-BEN-2024B08-060 and EHPC-EXT-2024E02-122 to access the EuroHPC supercomputer MareNostrum5 hosted the Barcelona Supercomputing Center (Spain).
} 

\backsection[Declaration of interests]{The authors report no conflict of interest.}
\backsection[Author ORCID]{
\\Marco De Paoli, \href{https://orcid.org/0000-0002-4709-4185}{https://orcid.org/0000-0002-4709-4185};
\\Guru Sreevanshu Yerragolam, \href{https://orcid.org/0000-0002-8928-2029}{https://orcid.org/0000-0002-8928-2029};
\\Roberto Verzicco \href{https://orcid.org/0000-0002-2690-9998}{https://orcid.org/0000-0002-2690-9998};
\\Detlef Lohse \href{https://orcid.org/0000-0003-4138-2255}{https://orcid.org/0000-0003-4138-2255}.
}

\bibliographystyle{jfm}
\bibliography{bibliography}

\end{document}